# AN EXPERIMENTAL AND KINETIC MODELLING STUDY OF THE OXIDATION OF THE FOUR ISOMERS OF BUTANOL

*Jeffrey T. MOSS, Andrew M. BERKOWITZ, Matthew A. OEHLSCHLAEGER*

Department of Mechanical, Aerospace, and Nuclear Engineering

Rensselaer Polytechnic Institute, Troy, NY, USA

*Joffrey BIET, Valérie WARTH, Pierre-Alexandre GLAUDE, Frédérique BATTIN-LECLERC*

Département de Chimie-Physique des Réactions, Nancy Université, CNRS, ENSIC, 1, rue Grandville, BP 20451, 54001 NANCY Cedex - France

## ABSTRACT

Butanol, an alcohol which can be produced from biomass sources, has received recent interest as an alternative to gasoline for use in spark ignition engines and as a possible blending compound with fossil diesel or biodiesel. Therefore, the autoignition of the four isomers of butanol (1-butanol, 2-butanol, iso-butanol, and tert-butanol) has been experimentally studied at high temperatures in a shock tube and a kinetic mechanism for description of their high-temperature oxidation has been developed. Ignition delay times for butanol/oxygen/argon mixtures have been measured behind reflected shock waves at temperatures and pressures ranging from approximately 1200 to 1800 K and 1 to 4 bar. Electronically excited OH emission and pressure measurements were used to determine ignition delay times. A detailed kinetic mechanism has been developed to describe the oxidation of the butanol isomers and validated by comparison to the shock tube measurements. Reaction flux and sensitivity analysis indicate that the consumption of 1-butanol and iso-butanol, the most reactive isomers, takes place primarily by H-atom abstraction resulting in the formation of radicals, the decomposition of which yields highly reactive branching agents, H-atoms and OH radicals. Conversely, the consumption of tert-butanol and 2-butanol, the least reactive isomers, takes place primarily via dehydration, resulting in the formation of alkenes, which lead to resonance stabilized radicals with very low reactivity. To our knowledge, the ignition delay measurements and oxidation mechanism presented here for 2-butanol, iso-butanol, and tert-butanol are the first of their kind.

## INTRODUCTION

Butanol derived from biomass sources, typically 1-butanol, has received recent interest as an alternative to gasoline and as a candidate for blending with diesel and biodiesel. Considerable research has been carried out aimed at production of butanol using fermentation (1)(2) and non-fermentative biosynthesis (3). It has been shown that butanol can be produced via ABE (acetone-butanol-ethanol) fermentation using starch- or sugar-based feedstocks (corn, wheat, sugar beets, sugar cane, sorghum) (4) and non-edible lignocellulosic biomass (5). Due to the possibilities of butanol as an alternative transportation fuel, recently the development and commercialization of butanol production processes has been undertaken by industrial corporations and startup companies. In partnership, Dupont and British Petroleum (BP) are converting a British Sugars ethanol plant in Wissington, England into a 1-butanol pilot plant to produce 9 million gallons yearly (6). The primary focus of synthesis research and commercial development has been on the production of 1-butanol, often called biobutanol, but there is also interest in processes to generate the higher octane rated 2-butanol and iso-butanol (2-methyl-1-propanol) isomers (3)(6) which may be used as neat fuels, in blended fuels, or as additives. Additionally, tert-butanol is currently used as a high octane rated (RON = 105, MON = 89) oxygenated additive in gasoline. However, tert-butanol has a melting point of 25.5 ºC and is not suitable as a neat fuel or as a high concentration blending compound because of its propensity to gel.

Butanol has several advantages over methanol and ethanol as a transportation fuel. The butanol isomers have higher energy density than methanol and ethanol, see Table I. The volumetric energy density of 1-butanol is only 9% lower than that for of gasoline. The butanols also have a much lower vapor pressure than ethanol and methanol reducing evaporative emissions and chance of explosion. They are less hygroscopic than ethanol and methanol making blending with gasoline and water contamination less problematic. 1-butanol has also shown to be less corrosive to materials found in automotive fuel systems and existing pipelines than ethanol and is very close in octane rating to gasoline (Table I). In fact, 1-butanol can be used as a direct replacement for gasoline in spark ignition engines with few or no modifications (7).

TABLE I

There have been few previous studies of the oxidation kinetics of the isomers of butanol. Rice *et al.* (8) and Yacoub *et al.* (9) have studied engine emissions and knock characteristics for a variety of butanol-gasoline blends. Hamins and Seshadri (10) investigated the extinction of butanol diffusion flames. McEnally and Pfefferle (11) studied methane/air flames doped with the four butanol isomers. They measured temperature, $C_1$-$C_{12}$ hydrocarbons, and major species allowing investigation of the decomposition kinetics of the butanol dopants and the emissions of toxic byproducts. Yang *et al.* (12) studied combustion intermediates in premixed, low-pressure (30 Torr, 1 Torr = 7.5 kPa), laminar, butanol-oxygen flames, for all four isomers, using photoionization mass spectrometry. Shock tube ignition delay measurements for 1-butanol have been performed by Zhukov *et al.* (13) for high-temperature, low-pressure (from 2.6 to 8.3 bar), argon-dilute mixtures. Dagaut and Togbé (14) have recently investigated the oxidation of blends of 1-butanol with a gasoline surrogate mixture (iso-octane, toluene, and 1-hexene blends) in jet-stirred reactor experiments (temperatures range from 770 to 1220 K at 10 bar). They also developed a kinetic oxidation mechanism for 1-butanol which was combined with a mechanism for their gasoline surrogate and used to simulate their jet-stirred reactor measurements with fairly good agreement between simulations and measurements. It has also been shown that the addition of oxygenated compounds, such as butanol, to traditional fuels can reduce engine soot emissions. In the recent study of Pepiot-Desjardins *et al.* (15) this phenomenon was quantified for a variety of oxygenates, including butanol.

Due to the growing interest in butanol as an alternative fuel and the limited number and types of previous kinetic studies, here we present shock tube ignition delay measurements for 1-butanol, 2-butanol, iso-butanol, and tert-butanol at high temperatures and a newly developed kinetic mechanism to describe the oxidation of the four isomers at high temperatures. To our knowledge, the measurements and mechanism presented here are the first of their kind for 2-butanol, iso-butanol, and tert-butanol.

## EXPERIMENTAL METHOD

Shock tube ignition delay times were measured in a newly constructed high-purity, low-pressure ($P < 10$ bar), stainless steel shock tube at Rensselaer Polytechnic Institute (RPI). The shock tube has a 12.3 cm inner diameter with 7.54 m long driven and 3.05 m long driver sections. The driver is separated from the driven by a single polycarbonate diaphragm prior to experiments. Diaphragms are burst using a stationary mechanical cutter by pressurizing the driver with helium. At the diaphragm section there is a round-square-round transition which allows the diaphragm to separate into four petals upon rupture, minimizing the formation of small diaphragm particles which can contaminate the test section.

The driven section is evacuated with a Varian DS202 roughing pump, which can evacuate the shock tube to $2\times10^{-3}$ Torr, and a Varian V70 turbomolecular pump coupled to a Varian DS202 backing pump, which can bring the shock tube to an ultimate pressure of $1\times10^{-6}$ Torr with a leak rate of $3\times10^{-6}$ Torr min$^{-1}$. Ultimate pressures are measured using an ion gauge (Varian type 564 gauge). The driver is evacuated with another Varian DS202 vacuum pump to ultimate pressures of $2\times10^{-3}$ Torr.

Reactant mixtures are made external to the shock tube in a stainless steel mixing vessel with an internal, magnetically-powered, vane stirrer. Gases are introduced to the mixing vessel from high-pressure gas cylinders (oxygen and argon) and via the vaporization of the liquid butanols from air-free glassware. The gases are introduced to the mixing vessel from the high-pressure cylinders and glassware through a stainless steel mixing manifold; after mixing the gas mixtures are introduced to the shock tube, located directly adjacent to the mixing vessel and manifold. The mixing vessel and manifold can be evacuated using a Varian DS202 vacuum pump to an ultimate pressure of $2\times10^{-3}$ Torr. Mixtures were made using the method of partial pressures with measurements of pressure made using a high accuracy Baratron pressure manometer and a Setra diaphragm pressure gauge. Reactant mixtures were made using high-purity oxygen (99.995%) and argon (99.999%) and 1-butanol (99.8+%), 2-butanol (99.5+%), iso-butanol (99.5+%), and tert-butanol (99.5+%) from Sigma Aldrich. The isomers of butanol were degassed prior to introduction into the evacuated mixing vessel via room-temperature evaporation. Atmospheric water contamination was not an issue with 1-butanol, 2-butanol, and iso-butanol which are only slightly miscible in water. On the other hand, tert-butanol, which is soluble in water, was frozen at ~10-20 ºC and pumped for a period of time to ensure that all atmospheric water introduced during transfer of the tert-butanol was removed. Mixtures were stirred using the magnetically-powered vane stirrer for two hours prior to experiments.

The shock tube is equipped with five piezoelectric pressure transducers (PCB transducer model 113A26 and amplifier model 498) with rise times of ≤1 μs spaced over the last 1.24 m of the driven section for the measurement of the incident shock velocity. The transducers are flush mounted in the shock tube sidewall and coated with room-temperature-vulcanizing (RTV) silicone insulation to avoid signal decay due to heat transfer to the transducer from the shock-heated gases. The transducers are equally separated by 30.5 cm with the last transducer located 2 cm from the endwall. The signals from the five pressure transducers are sent to four Raycal Dana 1992 universal counters (0.1 μs resolution) which provide the time intervals for incident shock passage allowing determination of the incident shock wave velocities at four locations over the last 1.1 m of the driven section. The four incident shock velocities are linearly extrapolated to the shock tube endwall to determine the incident shock velocity at the test location. Additionally, the pressure transducer located 2 cm from the shock tube endwall can be used for quantitative pressure measurements at the test location.

The incident and reflected shock conditions (vibrationally equilibrated) are calculated using the normal shock relations and the measured incident shock velocity at the test location, the measured initial temperature and pressure, and the thermodynamic properties of the reactant mixture. The initial temperature (room temperature) is measured using a mercury thermometer and the initial pressure is measured using the high accuracy Baratron pressure manometer located on the mixing/gas filling manifold directly adjacent to the shock tube. Thermodynamic properties of the mixtures are calculated using thermochemical polynomials from the Burcat and Ruscic database (16). The uncertainty in the resulting reflected shock conditions is estimated at <1 % and <1.5 % for temperature and pressure, respectively, based on uncertainties in the measured incident shock velocity and initial conditions (temperature, pressure, and mixture composition). The uncertainty in measured incident shock velocity is the largest contributor to uncertainty in reflected shock conditions. The measured reflected shock pressures agreed well with calculated pressures using vibrationally equilibrated incident and reflected shock conditions for all mixtures (±2%, within the uncertainty in measured reflected shock pressure).

Ignition delay times were determined behind reflected shock waves using the emission from electronically excited OH radicals (OH*) observed through a UV fused silica optic flush mounted in the shock tube endwall. The OH* emission was separated from other optical interference using a UG-5 Schott glass filter and recorded with a Thorlabs PDA36A silicon photodetector. The ignition delay time was defined as the time interval between shock arrival and reflection at the endwall and the onset of ignition at the endwall. The onset of ignition was determined by extrapolating the peak slope in OH* emission to the baseline and the shock arrival and reflection at the endwall was determined from the measured incident shock velocity and measurement of the shock passage at a location 2 cm from the endwall, where a piezoelectric transducer is located. See figure 1 for an example ignition time measurement. The pressure transducer and emission signals were recorded using a National Instruments PCI-6133 data acquisition card (3 MHz, 14-bit, and eight analog input channels) interfaced to a desktop computer with LabVIEW software.

FIGURE 1

Prior to performing measurements for butanol ignition delay times, measurements were made of ignition delay times for propane and methanol behind reflected shock waves at the conditions of previous shock tube studies published in the literature to validate the new shock tube facility and experimental techniques. Measurements for ignition delay times were made for propane/$O_2$/Ar mixtures at conditions previously investigated by Horning *et al.* (17) and measurements were made for methanol/$O_2$/Ar mixtures at conditions previously investigated by Shin *et al.* (18). These compounds were chosen due to the confidence we have in these two previous data sets and because methanol, as a liquid phase alcohol at room temperature, requires vaporization for mixture preparation, as do the butanol isomers. The validation measurements performed in the new shock tube are in excellent agreement (differences of at most 20%) with the previous studies lending confidence to the new facility and methods used for determination of reflected shock conditions and ignition times.

## EXPERIMENTAL RESULTS

Ignition delay measurements for 1-butanol, 2-butanol, iso-butanol, and tert-butanol were carried out at a variety of conditions to determine the ignition delay time dependence on concentration of the fuel, $O_2$, and argon and temperature. Measurements were made for all butanol isomers using mixtures (butanol/$O_2$/Ar molar %) with common composition at similar pressures: 1%/6%/93% ($\Phi$ = 1), 0.5%/3%/96.5% ($\Phi$ = 1), 1%/12%/87% ($\Phi$ = 0.5), 0.5%/6%/93.5% ($\Phi$ = 0.5), 1%/24%/75% ($\Phi$ = 0.25), and 0.5%/12%/87.5% ($\Phi$ = 0.25) mixtures all a pressures near 1 bar and 0.25%/1.5%/98.25% ($\Phi$ = 1) mixtures at pressures near 4 bar. Measurements were made for reflected shock temperatures ranging from 1196 to 1823 K with ignition times ranging from 47 to 1974 μs.

In figure 2 raw ignition delay results for 1-butanol are shown. As expected, the data exhibits clear Arrhenius exponential dependence on inverse temperature, with very little scatter about the linear least-square fits for given data sets. We estimate the uncertainty in ignition delay measurements at ±15% based on uncertainties in reflected shock temperature and pressure, initial mixture composition, and uncertainties in determining ignition times from the measured pressure and emission signals. The results for the three other isomers, not shown, have similar scatter to the 1-butanol data, with ignition times that vary in magnitude and slightly in overall activation energy. All of the data (177 experiments in total) is available in tabulated form in the supplemental data.

FIGURE 2

A linear regression analysis was performed to determine the ignition time dependence on reactant concentrations and temperature. The results of the regression analysis are four correlations for the butanol isomers:

*1-butanol*: $\tau = 10^{-8.34\,(\pm 0.47)}\,[\text{1-butanol}]^{-0.05\,(\pm 0.06)}\,[O_2]^{-0.50\,(\pm 0.04)}\,[Ar]^{-0.30\,(\pm 0.05)}\,\exp(18800 \pm 400 / T)$ (μs)

*2-butanol*: $\tau = 10^{-8.53\,(\pm 0.57)}\,[\text{2-butanol}]^{0.43\,(\pm 0.06)}\,[O_2]^{-1.12\,(\pm 0.04)}\,[Ar]^{-0.36\,(\pm 0.06)}\,\exp(18100 \pm 400 / T)$ (μs)

*iso-butanol*: $\tau = 10^{-10.19\,(\pm 0.62)}\,[\text{iso-butanol}]^{-0.04\,(\pm 0.06)}\,[O_2]^{-0.80\,(\pm 0.04)}\,[Ar]^{-0.34\,(\pm 0.05)}\,\exp(18800 \pm 450 / T)$ (μs)

*tert-butanol*: $\tau = 10^{-8.74(\pm 0.42)}\,[\text{tert-butanol}]^{0.22\,(\pm 0.05)}\,[O_2]^{-0.94\,(\pm 0.03)}\,[Ar]^{-0.19\,(\pm 0.04)}\,\exp(21300 \pm 400 / T)$ (μs)

where concentrations are in units of $mol/cm^3$, temperature is in Kelvin, and the given uncertainties in the correlation parameters are the uncertainties in the regression analysis resulting from scatter in the data. The correlated ignition times are shown in figure 3 and exhibit little deviation, with $r^2$ values of 98.7-99.5% for the four isomers. However, due to the relatively small range of conditions studied, extrapolating these correlations to conditions vastly different from those of the present study should be avoided.

FIGURE 3

The ignition times for the four butanol isomers are compared for a common mixture composition (1% butanol / 6% $O_2$ / 93% Ar, $\Phi = 1$) and pressure (~1 bar) in figure 4. The comparison shows that the ignition times vary from longest to shortest (least reactive to most reactive) in the order: tert-butanol, 2-butanol, iso-butanol, and 1-butanol. The figure also shows that the apparent activation energy is similar for 1-butanol, 2-butanol, and iso-butanol but is somewhat greater for tert-butanol; the greater activation energy for tert-butanol is also exhibited in the correlations.

FIGURE 4

To our knowledge, there has been only one previous study of butanol ignition. Zhukov *et al.* (13) recently reported shock tube ignition delay times for 1-butanol measured at pressures of 2.6, 4.4, and 8.3 bar and for three mixtures with $\Phi = 0.5$, 1.0, and 2.0. The majority of their measurements were made for $\Phi = 1.0$ mixtures with composition of 0.6% 1-butanol / 3.6% $O_2$ / 95.8% Ar. In figure 5 the current Φ = 1.0 1-butanol data is compared to the Φ = 1.0 1-butanol data of Zhukov *et al.*, with both data sets scaled to 1% 1-butanol / 6% $O_2$ / 93% Ar and 1 bar using the correlations given previously. When scaled, the agreement between the two studies is quite good.

FIGURE 5

## MECHANISM GENERATION

A single mechanism has been generated for the four isomers of butanol using the EXGAS software. This software has already been extensively described for the case of alkanes (19)(20)(21), as well as for ethers (22).

*General features of EXGAS*

The software provides reaction mechanisms consisting of three parts:

- A comprehensive primary mechanism, in which the only molecular reactants considered are the initial organic compounds and oxygen (see Table II).
- A lumped secondary mechanism, containing reactions that consume the molecular products of the primary mechanism which do not react in the reaction bases (19)(20).
- A $C_0$-$C_2$ reaction base including all the reactions involving radicals or molecules containing less than three carbon atoms (23), which is periodically updated and which is coupled with a reaction base for $C_3$-$C_4$ unsaturated hydrocarbons (24)(25), such as propyne, allene, butadiene or butenes, including reactions leading to the formation of benzene. The pressure-dependent rate constants follow the formalism proposed by Troe (26) and collision efficiency coefficients have been included.

Thermochemical data for molecules or radicals are automatically computed using the THERGAS software (27) based on group additivity (28) and stored as 14 polynomial coefficients, according to the CHEMKIN II formalism (29).

The rate parameters for isomerizations, combinations of free radicals, and unimolecular decompositions are calculated using the KINGAS software (19) and based on thermochemical kinetics methods (28), transition state theory, or modified collision theory. The kinetic data of other types of reactions are estimated from correlations (20), which are based on quantitative structure-reactivity relationships and obtained from the literature.

*Mechanism generation for alcohol reactants*

The structure of the current butanol mechanism is similar to that for alkanes except for the addition of a new generic dehydration reaction. Many rate constants were also modified for reactions involving bonds in the vicinity of the OH group. Table II presents the comprehensive primary mechanism, which contains 237 reactions for 1-butanol, 2-butanol, iso-butanol, tert-butanol and produced free radicals. Since the mechanism was built for modelling high-temperature conditions, the additions of alkyl free radicals to molecular oxygen are neglected. The generic reactions taken into account are then:

- Intramolecular dehydration (reactions 1-5 in Table II).

Intramolecular dehydration must be considered for alcohols and, therefore, has been added for the four butanols. Figure 6 illustrates the occurrence of this reaction via the favored four center cyclic transition state. Considering only the favored four center cyclic transition state, there is one possible intramolecular dehydration reaction for 1-butanol, iso-butanol and tert-butanol, and two for 2-butanol. Few kinetic parameters have been published concerning these reactions, e.g. by Tsang (30) in the case of ethanol ($A = 2.86 \times 10^{14}\ s^{-1}$ (3 abstractable H-atoms), Ea = 68.9 kcal/mol using a two parameters fit), by Bui *et al.* (31) from theoretical calculations for iso-propanol ($k_\infty = 2.0 \times 10^{6} T^{2.12} \exp(-30700/T)\ s^{-1}$ at the high pressure limit, for 6 abstractable H-atoms) and by Newman (32) for tert-butanol ($k = 7.7 \times 10^{14} \exp(-33145/T)\ s^{-1}$, for 9 abstractable H-atoms), but disagreements can be observed between all the proposed rate constants. For iso-butanol, we have used the iso-propanol value proposed by Bui *et al.* (31). For the other compounds, the rate parameters of these reactions which have large sensitivity coefficients, as it will be shown later in the text, have been fitted in order to obtain sufficient agreement with the experimental results. The starting point for this fit was taken from the recommendation of Tsang (30) for ethanol, taking into account the change in the number of abstractable H-atoms in the A-factor for the case of the butanols. The required maximum adjustment in the rate parameters, from this starting point, was a factor 5 for the A-factor (2-butanol) and adjustment to the activation energy of 3.1 kcal/mol (1-butanol and tert-butanol). The adjusted rate constant used for tert-butanol intramolecular dehydration is 3.75 times lower at 1200 K than that proposed by Newman (32). As shown by Tsang (30) in the case of the dehydration of ethanol, the fall-off effect starts to have some influence above 1800K; this effect has been neglected in the present work which concerns the heavier butanols as opposed to ethanol, the subject of Tsang's work. The formation of the different butene isomers, namely 1-butene, 2-butene and iso-butene, has been distinguished.

FIGURE 6

- ♦ Unimolecular initiations via C–C, C-O, or C–H bond fission (reactions 6-22).

The kinetic data for unimolecular decompositions were calculated using the KINGAS software (19) without considering fall-off effects. The calculation was made for a temperature of 1400K to reflect the range of experimental temperatures.

- ♦ Bimolecular initiations (reactions 23-38).

The rate constants for the bimolecular initiations between butanols and $O_2$ involving the abstraction of alkylic or alcoholic H-atoms were estimated using the correlation proposed by Ingham *et al.* (33) for hydrocarbons.

- ♦ Intramolecular radical isomerizations (reactions 39-46).

The isomerizations involving the transfer of an alcoholic H-atom were not considered. As in previous work (19)(20)(21), the activation energy was set equal to the sum of the activation energy for H-abstraction from the substrate by analogous radicals and the strain energy of the saturated cyclic transition state.

♦ Decompositions of radicals via β-scission (reactions 47-114).

In order to maintain a mechanism of manageable size, molecules formed in the primary mechanism, with the same molecular formula and the same functional groups, are lumped into one unique species without distinction for the different isomers. Consequently, the molecules obtained via β-scission and oxidation reactions are lumped. This is the case, for example, for butenols ($C_4H_8O$-LY in Table II), butanals ($C_4H_8O$-A), pentanals ($C_5H_{10}O$-A), pentanes ($C_5H_{12}$), and hexanes ($C_6H_{14}$). The molecules contained in the reaction bases are not lumped, e.g., allene, propyne and butenes. Kinetic data for reactions involving radical decomposition via β-scission are the same as those used in the case of alkanes and ethers (20)(22), apart from the activation energy of the β-scissions involving the breaking of a C-C or a C-H bond in α-position of the alcohol function. Activation energies for these reactions were evaluated using Evans-Polanyi correlations between the activation energies for alkyl and alkenyl free radical decomposition (20)(34) and enthalpies of reaction obtained using the THERGAS software (27).

♦ Oxidations (reactions 115-138).

The oxidations of the involved radicals by $O_2$ yield butenols or butanals. The rate constants are those used in previous work concerning alkanes (20) and alkenes (34).

♦ Metatheses involving an H-abstraction by small radicals (reactions 139-218).

The small radicals taken into account are O- and H-atoms and OH, $HO_2$ and $CH_3$. All the abstractions involving alkylic and alcoholic H-atoms have been considered. Kinetic data are generally those used for the alkanes (20). Rate constants for the abstraction of an H-atom from a carbon atom located in α-position of the alcohol function (reactions 139-153), the fastest of the H-atom abstraction reactions, have been evaluated using an Evans–Polanyi type correlation from Dean and Bozzelli (35), developed for the abstraction of H-atoms from hydrocarbons:

$$k = n_H A T^n \exp(-\{E_0 - f(\Delta H_0 - \Delta H)\}/RT)$$

where $n_H$ is the number of abstractable H-atoms; A, n, and $E_0$ are the rate parameters for the case of a metathesis by the considered radical from ethane; $\Delta H_0$ is the enthalpy of the metathesis by the considered radical from ethane; $\Delta H$ is the enthalpy of the metathesis by the considered radical from the reacting molecule; f is a correlation factor, the values of which are given by Dean and Bozzelli (35) for each considered radical; and R is the gas constant. According to Luo (36), the bond energy of the O-H bond for the alcohol function is between 102 and 106 $kcal.mol^{-1}$, a value equal to that of a C-H bond in the case of an alkylic primary H-atom. Therefore, the kinetic parameters for an H-atom abstraction from an alcohol function (reactions 154-173) are assumed equal to those for the abstraction of a primary alkylic H-atom.

- Termination steps (reactions 219-237).

Termination steps include the combinations involving iso-propyl and tert-butyl radicals with rate constants calculated using KINGAS software (19).

- Additional reactions.

Since butenes are important products of the oxidation of butanols under the present conditions, the mechanism for their consumption must be included. The reactions considered for the consumption of butenes (1-butene, 2-butene and iso-butene) were the additions of H- and O-atoms and of OH, $CH_3$ and $HO_2$ radicals to the butene double bonds and the H-abstractions by small radicals leading to resonance stabilized radicals. Additionally, two $C_2$ alcohol radicals, for which reactions were not considered in the $C_0$-$C_2$ reaction base, must been taken into account. The consumption of these two radicals, namely $^{\bullet}CH_2$-$CH_2$-OH and $CH_3$-$^{\bullet}CH$-OH, were taken from the mechanism proposed by Konnov (37).

TABLE II

## DISCUSSION

Simulations were performed using the SENKIN/CHEMKIN II software (29). Figures 7 to 11 display a comparison between the experimental ignition delay measurements and ignition delay times computed using the described mechanism, which globally contains 158 species and 1250 reactions; the complete mechanism is contained in the supplementary material. The reflected shock environment was simulated using a constant volume adiabatic constraint and the computed ignition times were defined as the time at which the OH concentration reached 10% of the maximum OH concentration occurring after ignition. The slight difference in experimental and computed ignition time definitions has a negligible influence on the results. The agreement between experiments and simulations is globally quite good, apart from disagreement for the more diluted mixtures at higher pressure (0.25% butanol, $\Phi$=1.0, and pressures between 3.5 and 4.3 bar; simulations not presented here) for which ignition delay times are underestimated by a up to a factor of three in the lower part of the studied temperature range. This disagreement is probably due to fall-off effects which have been neglected in the mechanism.

FIGURES 7 TO 11

The mechanism reproduces well the observed differences in reactivity for the four isomers under the different conditions studied as is illustrated in figure 7 for the case of stoichiometric mixtures containing 1% butanol at pressures around 1.2 bar. The most reactive species is 1-butanol; the reactivity of iso-butanol is just slightly lower. Longer ignition times were observed and predicted via simulation for 2-butanol and the least reactive isomer tert-butanol, for which the measured and simulated ignition times have a larger apparent activation energy compared to the three other isomers.

Figures 8a, 9a, 10a and 11a exhibit that the experimentally observed increase in ignition delay time as the equivalence ratio varies from 0.25 to 1.0 at pressures around 1.2 bar is mostly well represented by the mechanism. Figures 8b, 9b, 10b and 11b illustrate that the experimentally observed acceleration in ignition with the increase in butanol concentration from 0.5 to 1.0% is captured by the simulations for the case of mixtures with an equivalence ratio of 0.5 at pressures around 1.2 bar. However the influence of equivalence ratio and initial concentration on ignition times is stronger in the simulations than experimentally observed, especially in the case of linear butanols for mixtures with the highest butanol concentrations.

Figure 12 displays a reaction flux rate analysis performed at 1450 K, for an equivalence ratio of 1.0, at atmospheric pressure, for mixtures containing 1% butanol, and for 50% conversion of butanol. Figures 13 and 14 present the temporal evolution of some major species and a sensitivity analysis, respectively computed under the same conditions.

FIGURES 12 to 14

1-butanol and iso-butanol are mainly consumed by H-abstractions by hydrogen atoms and hydroxyl radicals yielding radicals, the decomposition of which leads to the formation of highly reactive radicals, such as the branching agents, H-atoms and OH radicals. The difference in reactivity between 1-butanol and iso-butanol is due to the fact that 69% of the consumption of 1-butanol leads to H-atoms, while only 29% in the case of iso-butanol, with 27% of its consumption leading to the less reactive methyl radical. As shown in figures 13a and 13c, the auto-ignition of 1-butanol and iso-butanol occurs not too far after the total consumption of the reactant. The primary products obtained from 1-butanol are butanal, acetaldehyde, and ethylene (not shown in the figure), and to a lesser extent propene and 1-butene. The consumption of iso-butanol leads to similar amounts of propenol, iso-butanal, and iso-butene.

The reaction pathways are very different for 2-butanol and tert-butanol which react primarily (almost 70% of their consumption) by dehydration to form alkenes which are consumed by additions of H-atoms or by H-abstractions, reactions with very small flux, to give very non-reactive resonance stabilized radicals. Iso-butene obtained from tert-butanol by dehydration, as well a by H-abstraction followed by a β-scission decomposition, yields only the very stable iso-butyl radical by H-abstraction, explaining the low reactivity of this branched alcohol. As shown in figures 13b and 13d, the autoignition of 2-butanol and tert-butanol occurs long after the total consumption of the reactant and the main primary products are alkenes, 1-butene and iso-butene, respectively. The autoignition occurs only when the alkenes are totally consumed. Oxygenated species, such as butanone for 2-butanol and propanone for tert-butanol are produced only in much smaller concentrations.

The higher apparent activation energy (ignition time slope) observed experimentally and predicted in the simulations for tert-butanol relative to the other isomers is due to the higher activation energies of all of the initiation reactions (dehydration, unimolecular decomposition, and H-abstraction) for tert-butanol relative to the other isomers and the higher activation energies of the iso-butene consumption reactions relative to the consumption reactions for the primary products of the other butanol isomers (i.e., the relative stability of iso-butene compared to the other primary products).

The sensitivity analysis in figure 14 shows that dehydration reactions for butanol have a large impact on the ignition delay times; the inhibiting effect of dehydration is particularly important in the case of 2-butanol. While the amount of butanol consumed through unimolecular initiations is relatively small (less than 10% of the butanol consumption for all of the isomers, as shown in figure 12), these reactions, which are a source of radicals, have a promoting effect, especially in the case of the less reactive species (2-butanol and tert-butanol). Metatheses reactions, which consume reactive species such H-atoms and OH radicals, have a slight inhibiting effect, apart from the case of 2-butanol for which metatheses have a promoting effect as they are the main channels competing with dehydration.

**CONCLUSIONS**

Ignition delay time measurements have been made and a kinetic oxidation mechanism has been developed for the four butanol isomers: 1-butanol, 2-butanol, iso-butanol, and tert-butanol. The measurements and kinetic mechanism are the first of their kind for 2-butanol, iso-butanol, and tert-butanol. The mechanism predictions provide good agreement for the experimentally observed differences in reactivity of the four butanol isomers. Reaction flux and sensitivity analysis illustrates the relative importance of the three classes of butanol consumption reactions: dehydration, unimolecular decomposition, and H-atom abstraction. Uncertainties remain for the rate constants of dehydration reactions, which have the major influence on the reactivity. The less reactive butanols, tert-butanol and 2-butanol, are those for which the formation of alkenes through dehydration reaction is preponderant. Additional data in shock tubes at higher pressures would be of interest to give more information on fall-off effects, as well as experiments in another type of apparatus involving analyses of the obtained products.

**ACKNOWLEDGEMENTS**

The RPI group thanks Peter Catalfamo, Patrick McComb, and Benjamin Waxman for assistance in performing the experiments and acknowledges the donors of the American Chemical Society Petroleum Research Fund for partial support of this research.

**SUPPORTING INFORMATION AVAILABLE:** The experimental data in tabular form and kinetic mechanism in CHEMKIN format are available free of charge at http://pubs.acs.org.

**TABLE I: Properties of gasoline, methanol, ethanol, and 1-butanol.**

| Fuel | Lower heating value [MJ/kg] | Volumetric energy density [MJ/L] | Research octane number (RON) | Motor octane number (MON) | Vapor pressure @ 25 ºC [Torr] |
|---|---|---|---|---|---|
| Gasoline | 42.5 | 32 | 92-98 | 82-88 | |
| Methanol | 19.9 | 16 | 136 | 104 | 127 |
| Ethanol | 28.9 | 20 | 129 | 102 | 59 |
| 1-butanol | 33.1 | 29 | 96 | 78 | 6 |

**TABLE II: Primary mechanism for the oxidation of the four butanol isomers at high temperature.**

The rate constants are given at 1 bar ($k = A\,T^n \exp(-E_a/RT)$) in $cm^3$, mol, s, cal units.

| Reactions | A | n | $E_a$ | n° |
|---|---|---|---|---|
| **Intramolecular Dehydrations :** | | | | |
| $C_4H_9OH$-1=>$H_2O$+$C_4H_8$-1 | $2.0.10^{+14}$ | 0.0 | 72000 | (1) |
| $C_4H_9OH$-2=>$H_2O$+$C_4H_8$-1 | $1.5.10^{+15}$ | 0.0 | 66000 | (2) |
| $C_4H_9OH$-2=>$H_2O$+$C_4H_8$-2 | $2.0.10^{+14}$ | 0.0 | 67000 | (3) |
| iso$C_4H_9OH$=>$H_2O$+i$C_4H_8$ | $2.0.10^{+6}$ | 2.12 | 62000 | (4) |
| ter$C_4H_9OH$=>$H_2O$+i$C_4H_8$ | $2.7.10^{+15}$ | 0.0 | 72000 | (5) |
| **Unimolecular Initiations :** | | | | |
| $C_4H_9OH$-1=•OH+•$C_4H_9$-1 | $1.16.10^{+15}$ | 0.0 | 92320 | (6) |
| $C_4H_9OH$-1=•H+$C_3H_7CH_2O$• | $1.51.10^{+13}$ | 0.0 | 103620 | (7) |
| $C_4H_9OH$-1=•$C_3H_7$-1+•$CH_2OH$ | $1.47.10^{+15}$ | 0.0 | 82930 | (8) |
| $C_4H_9OH$-1=•$C_2H_5$+•$CH_2CH_2OH$ | $2.23.10^{+15}$ | 0.0 | 82760 | (9) |
| $C_4H_9OH$-1=•$CH_3$+•$CH_2C_2H_4OH$ | $5.82.10^{+15}$ | 0.0 | 84870 | (10) |
| $C_4H_9OH$-2=•$C_2H_5$•CHOH+$CH_3$ | $1.58.10^{+15}$ | 0.0 | 79770 | (11) |
| $C_4H_9OH$-2=•OH+•$C_4H_9$-2 | $3.25.10^{+14}$ | 0.0 | 92050 | (12) |
| $C_4H_9OH$-2=•H+$C_2H_5$CH(O•)$CH_3$ | $1.24.10^{+13}$ | 0.0 | 105460 | (13) |
| $C_4H_9OH$-2=•$C_2H_5$+$CH_3$•CHOH | $6.06.10^{+14}$ | 0.0 | 82170 | (14) |
| $C_4H_9OH$-2=•$CH_3$+•$CH_2$CH(OH)$CH_3$ | $1.83.10^{+15}$ | 0.0 | 85370 | (15) |
| iso$C_4H_9OH$=•$CH_3$+$CH_3$•$CHCH_2OH$ | $5.14.10^{+15}$ | 0.0 | 80540 | (16) |
| iso$C_4H_9OH$=•$CH_2OH$+•$C_3H_7$-2 | $6.47.10^{+14}$ | 0.0 | 79070 | (17) |
| iso$C_4H_9OH$=•OH+•i$C_4H_9$ | $1.40.10^{+14}$ | 0.0 | 91720 | (18) |
| iso$C_4H_9OH$=•H+•$CH_3CH(CH_3)CH_2O$• | $5.24.10^{+12}$ | 0.0 | 101860 | (19) |
| ter$C_4H_9OH$=•OH+•t$C_4H_9$ | $1.00.10^{+15}$ | 0.0 | 95639 | (20) |
| ter$C_4H_9OH$=•H+$C(CH_3)_3O$• | $2.39.10^{+13}$ | 0.0 | 105630 | (21) |
| ter$C_4H_9OH$=•$CH_3$+•$C(CH_3)_3OH$ | $1.9.10^{+16}$ | 0.0 | 83600 | (22) |
| **Bimolecular Initiations :** | | | | |
| $C_4H_9OH$-1+$O_2$=•$HO_2$+$C_3H_7$•CHOH | $1.4.10^{+13}$ | 0.0 | 46869 | (23) |
| $C_4H_9OH$-1+$O_2$=•$HO_2$+$C_3H_7CH_2O$• | $7.0.10^{+12}$ | 0.0 | 55172 | (24) |
| $C_4H_9OH$-1+$O_2$=•$HO_2$+•$CH_2C_3H_6OH$ | $2.1.10^{+13}$ | 0.0 | 53033 | (25) |
| $C_4H_9OH$-1+$O_2$=•$HO_2$+$CH_3$•$CHC_2H_4OH$ | $1.4.10^{+13}$ | 0.0 | 50588 | (26) |
| $C_4H_9OH$-1+$O_2$=•$HO_2$+$C_2H_5$•$CHCH_2OH$ | $1.4.10^{+13}$ | 0.0 | 50652 | (27) |
| $C_4H_9OH$-2+$O_2$=•$HO_2$+$C_2H_5$CH(OH)•$CH_2$ | $2.1.10^{+13}$ | 0.0 | 52333 | (28) |
| $C_4H_9OH$-2+$O_2$=•$HO_2$+•$CH_2CH_2$CH(OH)$CH_3$ | $2.1.10^{+13}$ | 0.0 | 53033 | (29) |
| $C_4H_9OH$-2+$O_2$=•$HO_2$+$CH_3$•CHCH(OH)CH3 | $1.4.10^{+13}$ | 0.0 | 50588 | (30) |
| $C_4H_9OH$-2+$O_2$=•$HO_2$+$C_2H_5$CH(O•)$CH_3$ | $7.0.10^{+12}$ | 0.0 | 57272 | (31) |
| $C_4H_9OH$-2+$O_2$=•$HO_2$+$C_2H_5$C•(OH)$CH_3$ | $7.0.10^{+12}$ | 0.0 | 44726 | (32) |
| iso$C_4H_9OH$+$O_2$=•$HO_2$+$CH_3$C• $(CH_3)CH_2OH$ | $7.0.10^{+12}$ | 0.0 | 47243 | (33) |
| iso$C_4H_9OH$+$O_2$=•$HO_2$+•$CH_2CH(CH_3)CH_2OH$ | $4.2.10^{+13}$ | 0.0 | 52333 | (34) |
| iso$C_4H_9OH$+$O_2$=•$HO_2$+$CH_3CH(CH_3)CH_2O$• | $7.0.10^{+12}$ | 0.0 | 55172 | (35) |
| iso$C_4H_9OH$+$O_2$=•$HO_2$+$CH_3CH(CH_3)$ •CHOH | $1.4.10^{+13}$ | 0.0 | 46869 | (36) |
| ter$C_4H_9OH$+$O_2$=•$HO_2$+$C(CH_3)_3O$• | $7.0.10^{+12}$ | 0.0 | 56872 | (37) |
| ter$C_4H_9OH$+$O_2$=•$HO_2$+•$CH_2C(CH_3)_2OH$ | $6.3.10^{+13}$ | 0.0 | 53033 | (38) |
| **Isomerizations :** | | | | |
| •$C_4H_9$-1=•$C_4H_9$-2 | $3.3.10^{+09}$ | 1.0 | 37000 | (39) |
| $C_3H_7$•CHOH=$CH_3$•$CHC_2H_4OH$ | $3.3.10^{+09}$ | 1.0 | 43000 | (40) |
| $C_3H_7$•CHOH=•$CH_2C_3H_6OH$ | $8.6.10^{+08}$ | 1.0 | 25800 | (41) |
| •$CH_2C_3H_6OH$=$C_2H_5$•$CHCH_2OH$ | $3.3.10^{+09}$ | 1.0 | 37000 | (42) |
| $C_2H_5$CH(OH)•$CH_2$=$CH_3$•CHCH(OH)$CH_3$ | $3.3.10^{+09}$ | 1.0 | 37000 | (43) |
| $C_2H_5CH(OH)\bullet CH_2$=$\bullet CH_2CH_2CH(OH)CH_3$ | $8.6.10^{+08}$ | 1.0 | 19800 | (44) |
| $\bullet CH_2CH_2CH(OH)CH_3$=$C_2H_5C\bullet$ $(OH)CH_3$ | $1.7.10^{+09}$ | 1.0 | 33000 | (45) |
| $\bullet CH_2CH(CH_3)CH_2OH$=$CH_3CH(CH_3)$ $\bullet CHOH$ | $3.3.10^{+09}$ | 1.0 | 35000 | (46) |
| **Beta-scissions :** | | | | |
| $\bullet C_3H_7$-1=>$\bullet CH_3$+$C_2H_4$ | $2.0.10^{+13}$ | 0.0 | 31000 | (47) |
| $\bullet C_3H_7$-1=>$\bullet H$+$C_3H_6$ | $3.0.10^{+13}$ | 0.0 | 38000 | (48) |
| $\bullet C_4H_9$-1=>$\bullet C_2H_5$+$C_2H_4$ | $2.0.10^{+13}$ | 0.0 | 28700 | (49) |
| $\bullet C_4H_9$-1=>$\bullet H$+$C_4H_8$-1 | $3.0.10^{+13}$ | 0.0 | 38000 | (50) |
| $C_3H_7\bullet CHOH$=>$\bullet H$+$C_4H_8O$-LY [a,b] | $3.0.10^{+13}$ | 0.0 | 36400 | (51) |
| $C_3H_7\bullet CHOH$=>$\bullet H$+$C_4H_8O$-A [c] | $2.5.10^{+13}$ | 0.0 | 29000 | (52) |
| $C_3H_7CH_2O\bullet$=>$\bullet C_3H_7$-1+HCHO | $2.0.10^{+13}$ | 0.0 | 24700 | (53) |
| $C_3H_7CH_2O\bullet$=>$\bullet H$+$C_4H_8O$-A | $3.0.10^{+13}$ | 0.0 | 27800 | (54) |
| $\bullet CH_2C_3H_6OH$=>$\bullet CH_2CH_2OH$+$C_2H_4$ | $2.0.10^{+13}$ | 0.0 | 30700 | (55) |
| $\bullet CH_2C_3H_6OH$=>$\bullet H$+$C_4H_8O$-LY | $3.0.10^{+13}$ | 0.0 | 36500 | (56) |
| $CH_3\bullet CHC_2H_4OH$=>$\bullet CH_2OH$+$C_3H_6$ | $2.0.10^{+13}$ | 0.0 | 30500 | (57) |
| $CH_3\bullet CHC_2H_4OH$=>$\bullet H$+$C_4H_8O$-LY | $3.0.10^{+13}$ | 0.0 | 36900 | (58) |
| $CH_3\bullet CHC_2H_4OH$=>$\bullet H$+$C_4H_8O$-LY | $3.0.10^{+13}$ | 0.0 | 38900 | (59) |
| $C_2H_5\bullet CHCH_2OH$=>$\bullet OH$+$C_4H_8$-1 | $2.0.10^{+13}$ | 0.0 | 26000 | (60) |
| $C_2H_5\bullet CHCH_2OH$=>$\bullet H$+$C_4H_8O$-LY | $3.0.10^{+13}$ | 0.0 | 36900 | (61) |
| $C_2H_5\bullet CHCH_2OH$=>$\bullet H$+$C_4H_8O$-LY | $3.0.10^{+13}$ | 0.0 | 31200 | (62) |
| $C_2H_5\bullet CHCH_2OH$=>$\bullet CH_3$+$C_3H_6O$-LY | $2.0.10^{+13}$ | 0.0 | 30600 | (63) |
| $C_2H_5CH(OH)\bullet CH_2$=>$\bullet OH$+$C_4H_8$-1 | $2.0.10^{+13}$ | 0.0 | 26000 | (64) |
| $C_2H_5CH(OH)\bullet CH_2$=>$\bullet H$+$C_4H_8O$-LY | $1.5.10^{+13}$ | 0.0 | 35000 | (65) |
| $\bullet CH_2CH_2CH(OH)CH_3$=>$CH_3\bullet CHOH$+$C_2H_4$ | $2.0.10^{+13}$ | 0.0 | 29600 | (66) |
| $\bullet CH_2CH_2CH(OH)CH_3$=>$\bullet H$+$C_4H_8O$-LY | $3.0.10^{+13}$ | 0.0 | 36900 | (67) |
| $CH_3\bullet CHCH(OH)CH_3$=>$\bullet OH$+$C_4H_8$-2 | $2.0.10^{+13}$ | 0.0 | 26000 | (68) |
| $CH_3\bullet CHCH(OH)CH_3$=>$\bullet H$+$C_4H_8O$-LY | $3.0.10^{+13}$ | 0.0 | 38100 | (69) |
| $CH_3\bullet CHCH(OH)CH_3$=>$\bullet CH_3$+$C_3H_6O$-LY | $2.0.10^{+13}$ | 0.0 | 29700 | (70) |
| $CH_3\bullet CHCH(OH)CH_3$=>$\bullet H$+$C_4H_8O$-LY | $1.5.10^{+13}$ | 0.0 | 34800 | (71) |
| $C_2H_5CH(O\bullet)CH_3$=>$\bullet C_2H_5$+$CH_3CHO$ | $2.0.10^{+13}$ | 0.0 | 21700 | (72) |
| $C_2H_5CH(O\bullet)CH_3$=>$\bullet CH_3$+$C_2H_5CHO$ | $2.0.10^{+13}$ | 0.0 | 21900 | (73) |
| $C_2H_5CH(O\bullet)CH_3$=>$\bullet H$+$C_3H_8CO$ | $1.5.10^{+13}$ | 0.0 | 25000 | (74) |
| $C_2H_5C\bullet(OH)CH_3$=>$\bullet H$+$C_3H_8CO$ | $2.5.10^{+13}$ | 0.0 | 29000 | (75) |
| $C_2H_5C\bullet(OH)CH_3$=>$\bullet H$+$C_4H_8O$-LY | $3.0.10^{+13}$ | 0.0 | 37800 | (76) |
| $C_2H_5C\bullet(OH)CH_3$=>$\bullet H$+$C_4H_8O$-LY | $3.0.10^{+13}$ | 0.0 | 39200 | (77) |
| $C_2H_5C\bullet(OH)CH_3$=>$\bullet CH_3$+$C_3H_6O$-LY | $2.0.10^{+13}$ | 0.0 | 32400 | (78) |
| $CH_3C\bullet(CH_3)CH_2OH$=>$\bullet OH$+$iC_4H_8$ | $2.0.10^{+13}$ | 0.0 | 26000 | (79) |
| $CH_3C\bullet(CH_3)CH_2OH$=>$\bullet H$+$C_4H_8O$-LY | $3.0.10^{+13}$ | 0.0 | 39100 | (80) |
| $CH_3C\bullet(CH_3)CH_2OH$=>$\bullet H$+$C_4H_8O$-LY | $6.0.10^{+13}$ | 0.0 | 35100 | (81) |
| $\bullet CH_2CH(CH_3)CH_2OH$=>$\bullet CH_2OH$+$C_3H_6$ | $2.0.10^{+13}$ | 0.0 | 30600 | (82) |
| $\bullet CH_2CH(CH_3)CH_2OH$=>$\bullet CH_3$+$C_3H_6O$-LY | $2.0.10^{+13}$ | 0.0 | 30600 | (83) |
| $\bullet CH_2CH(CH_3)CH_2OH$=>$\bullet H$+$C_4H_8O$-LY | $1.5.10^{+13}$ | 0.0 | 36700 | (84) |
| $CH_3CH(CH_3)CH_2O\bullet$=>$\bullet C_3H_7$-2+HCHO | $2.0.10^{+13}$ | 0.0 | 24200 | (85) |
| $CH_3CH(CH_3)CH_2O\bullet$=>$\bullet H$+$C_4H_8O$-A | $3.0.10^{+13}$ | 0.0 | 28600 | (86) |
| $CH_3CH(CH_3)\bullet CHOH$=>$\bullet H$+$C_4H_8O$-LY | $3.0.10^{+13}$ | 0.0 | 35600 | (87) |
| $CH_3CH(CH_3)\bullet CHOH$=>$\bullet H$+$C_4H_8O$-A | $2.5.10^{+13}$ | 0.0 | 29000 | (88) |
| $CH_3CH(CH_3)\bullet CHOH$=>$\bullet CH_3$+$C_3H_6O$-LY | $4.0.10^{+13}$ | 0.0 | 30500 | (89) |
| $C(CH_3)_3O\bullet$=>$\bullet CH_3$+$C_2H_6CO$ | $6.0.10^{+13}$ | 0.0 | 22700 | (90) |
| $\bullet CH_2C(CH_3)_2OH$=>$\bullet OH$+$iC_4H_8$ | $4.0.10^{+13}$ | 0.0 | 26000 | (91) |
| $\bullet CH_2C(CH_3)_2OH$=>$\bullet CH_3$+$C_3H_6O$-LY | $4.0.10^{+13}$ | 0.0 | 32500 | (92) |
| $\bullet C_4H_9$-2=>$\bullet CH_3$+$C_3H_6$ | $2.0.10^{+13}$ | 0.0 | 31000 | (93) |
| $\bullet C_4H_9$-2=>$\bullet H$+$C_4H_8$-1 | $3.0.10^{+13}$ | 0.0 | 38000 | (94) |
| $\bullet C_4H_9$-2=>$\bullet H$+$C_4H_8$-2 | $3.0.10^{+13}$ | 0.0 | 39000 | (95) |
| $\bullet C_3H_7$-2=>$\bullet H$+$C_3H_6$ | $6.0.10^{+13}$ | 0.0 | 39000 | (96) |
| $\bullet CH_2C_2H_4OH$=>$\bullet CH_2OH$+$C_2H_4$ | $2.0.10^{+13}$ | 0.0 | 31100 | (97) |
| $\bullet CH_2C_2H_4OH$=>$\bullet H$+$C_3H_6O$-LY | $3.0.10^{+13}$ | 0.0 | 36500 | (98) |
| $C_2H_5\bullet CHOH$=>$\bullet H$+$C_2H_5CHO$ | $2.5.10^{+13}$ | 0.0 | 29000 | (99) |
| $C_2H_5\bullet CHOH$=>$\bullet H$+$C_3H_6O$-LY | $3.0.10^{+13}$ | 0.0 | 36800 | (100) |
| $\bullet CH_2CH(OH)CH_3$=>$\bullet OH$+$C_3H_6$ | $2.0.10^{+13}$ | 0.0 | 26000 | (101) |
| $\bullet CH_2CH(OH)CH_3$=>$\bullet H$+$C_3H_6O$-LY | $1.5.10^{+13}$ | 0.0 | 35200 | (102) |
| $CH_3\bullet CHCH_2OH$=>$\bullet OH$+$C_3H_6$ | $2.0.10^{+13}$ | 0.0 | 26000 | (103) |
| $CH_3\bullet CHCH_2OH$=>$\bullet H$+$C_3H_6O$-LY | $3.0.10^{+13}$ | 0.0 | 37700 | (104) |
| $CH_3\bullet CHCH_2OH$=>$\bullet H$+$C_3H_6O$-LY | $3.0.10^{+13}$ | 0.0 | 34700 | (105) |
| $\bullet iC_4H_9=>\bullet CH_3+C_3H_6$ | $4.0.10^{+13}$ | 0.0 | 31000 | (106) |
| $\bullet iC_4H_9=>\bullet H+iC_4H_8$ | $3.0.10^{+13}$ | 0.0 | 37500 | (107) |
| $\bullet tC_4H_9=>\bullet H+iC_4H_8$ | $9.0.10^{+13}$ | 0.0 | 39000 | (108) |
| $\bullet C(CH_3)_3OH=>\bullet H+C_2H_6CO$ | $2.5.10^{+13}$ | 0.0 | 29000 | (109) |
| $\bullet C(CH_3)_3OH=>\bullet H+C_3H_6O$-LY | $6.0.10^{+13}$ | 0.0 | 39400 | (110) |
| $C_3H_7\bullet CHOH=>CH_3CHO+\bullet C_2H_5$ | $2.0.10^{+13}$ | 0.0 | 28700 | (111) |
| $C_2H_5CH(OH)\bullet CH_2=>CH_3CHO+\bullet C_2H_5$ | $2.0.10^{+13}$ | 0.0 | 28700 | (112) |
| $C_2H_5\bullet CHOH=>CH_3CHO+\bullet CH_3$ | $2.0.10^{+13}$ | 0.0 | 31000 | (113) |
| $\bullet CH_2CH(OH)CH_3=>CH_3CHO+\bullet CH_3$ | $2.0.10^{+13}$ | 0.0 | 31000 | (114) |
| **Oxidations :** | | | | |
| $C_3H_7\bullet CHOH+O_2=>C_4H_8O$-LY$+\bullet HO_2$ | $1.3.10^{+12}$ | 0.0 | 5000 | (115) |
| $C_3H_7\bullet CHOH+O_2=>C_4H_8O$-A$+\bullet HO_2$ | $4.4.10^{+11}$ | 0.0 | 5000 | (116) |
| $C_3H_7CH_2O\bullet+O_2=>C_4H_8O$-A$+\bullet HO_2$ | $1.3.10^{+12}$ | 0.0 | 5000 | (117) |
| $\bullet CH_2C_3H_6OH+O_2=>C_4H_8O$-LY$+\bullet HO_2$ | $1.3.10^{+12}$ | 0.0 | 5000 | (118) |
| $CH_3\bullet CHC_2H_4OH+O_2=>C_4H_8O$-LY$+\bullet HO_2$ | $1.3.10^{+12}$ | 0.0 | 5000 | (119) |
| $CH_3\bullet CHC_2H_4OH+O_2=>C_4H_8O$-LY$+\bullet HO_2$ | $5.3.10^{+11}$ | 0.0 | 5000 | (120) |
| $C_2H_5\bullet CHCH_2OH+O_2=>C_4H_8O$-LY$+\bullet HO_2$ | $1.3.10^{+12}$ | 0.0 | 5000 | (121) |
| $C_2H_5\bullet CHCH_2OH+O_2=>C_4H_8O$-LY$+\bullet HO_2$ | $1.3.10^{+12}$ | 0.0 | 5000 | (122) |
| $C_2H_5CH(OH)\bullet CH_2+O_2=>C_4H_8O$-LY$+\bullet HO_2$ | $4.4.10^{+11}$ | 0.0 | 5000 | (123) |
| $\bullet CH_2CH_2CH(OH)CH_3+O_2=>C_4H_8O$-LY$+\bullet HO_2$ | $1.3.10^{+12}$ | 0.0 | 5000 | (124) |
| $CH_3\bullet CHCH(OH)CH_3+O_2=>C_4H_8O$-LY$+\bullet HO_2$ | $4.4.10^{+11}$ | 0.0 | 5000 | (125) |
| $CH_3\bullet CHCH(OH)CH_3+O_2=>C_4H_8O$-LY$+\bullet HO_2$ | $5.3.10^{+11}$ | 0.0 | 5000 | (126) |
| $C_2H_5CH(O\bullet)CH_3+O_2=>C_3H_8CO+\bullet HO_2$ | $4.4.10^{+11}$ | 0.0 | 5000 | (127) |
| $C_2H_5C\bullet(OH)CH_3+O_2=>C_4H_8O$-LY$+\bullet HO_2$ | $1.3.10^{+12}$ | 0.0 | 5000 | (128) |
| $C_2H_5C\bullet(OH)CH_3+O_2=>C_4H_8O$-LY$+\bullet HO_2$ | $5.3.10^{+11}$ | 0.0 | 5000 | (129) |
| $C_2H_5C\bullet(OH)CH_3+O_2=>C_3H_8CO+\bullet HO_2$ | $4.4.10^{+11}$ | 0.0 | 5000 | (130) |
| $CH_3C\bullet(CH_3)CH_2OH+O_2=>C_4H_8O$-LY$+\bullet HO_2$ | $1.3.10^{+12}$ | 0.0 | 5000 | (131) |
| $CH_3C\bullet(CH_3)CH_2OH+O_2=>C_4H_8O$-LY$+\bullet HO2$ | $1.0.10^{+12}$ | 0.0 | 5000 | (132) |
| $\bullet CH_2CH(CH_3)CH_2OH+O_2=>C_4H_8O$-LY$+\bullet HO_2$ | $4.4.10^{+11}$ | 0.0 | 5000 | (133) |
| $CH_3CH(CH_3)CH_2O\bullet+O_2=>C_4H_8O$-A$+\bullet HO_2$ | $1.3.10^{+12}$ | 0.0 | 5000 | (134) |
| $CH_3CH(CH_3)\bullet CHOH+O_2=>C_4H_8O$-LY$+\bullet HO_2$ | $4.4.10^{+11}$ | 0.0 | 5000 | (135) |
| $CH_3CH(CH_3)\bullet CHOH+O_2=>C_4H_8O$-A$+\bullet HO_2$ | $4.4.10^{+11}$ | 0.0 | 5000 | (136) |
| $\bullet C_3H_7$-2$+O_2=>C_3H_6+\bullet HO_2$ | $2.3.10^{+12}$ | 0.0 | 5000 | (137) |
| $\bullet tC_4H_9+O_2=>iC_4H_8+\bullet HO_2$ | $1.6.10^{+12}$ | 0.0 | 5000 | (138) |
| **Metathesis :** | | | | |
| $C_4H_9OH$-1$+\bullet O\bullet=>\bullet OH+C_3H_7\bullet CHOH$ | $3.4.10^{+08}$ | 1.5 | 1000 | (139) |
| $C_4H_9OH$-1$+\bullet H=>H_2+C_3H_7\bullet CHOH$ | $4.8.10^{+09}$ | 1.5 | 3310 | (140) |
| $C_4H_9OH$-1$+\bullet OH=>H_2O+C_3H_7\bullet CHOH$ | $2.4.10^{+06}$ | 2.0 | -2200 | (141) |
| $C_4H_9OH$-1$+\bullet HO_2=>H_2O_2+C_3H_7\bullet CHOH$ | $2.8.10^{+04}$ | 2.7 | 14380 | (142) |
| $C_4H_9OH$-1$+\bullet CH_3=>CH_4+C_3H_7\bullet CHOH$ | $1.6.10^{+06}$ | 1.9 | 6840 | (143) |
| $C_4H_9OH$-2$+\bullet O\bullet=>\bullet OH+C_2H_5C\bullet(OH)CH_3$ | $1.7.10^{+08}$ | 1.5 | -350 | (144) |
| $C_4H_9OH$-2$+\bullet H=>H_2+C_2H_5C\bullet(OH)CH_3$ | $2.4.10^{+09}$ | 1.5 | 2140 | (145) |
| $C_4H_9OH$-2$+\bullet OH=>H_2O+C_2H_5C\bullet(OH)CH_3$ | $1.2.10^{+06}$ | 2.0 | -3100 | (146) |
| $C_4H_9OH$-2$+\bullet HO_2=>H_2O_2+C_2H_5C\bullet(OH)CH_3$ | $1.4.10^{+04}$ | 2.7 | 13300 | (147) |
| $C_4H_9OH$-2$+\bullet CH_3=>CH_4+C_2H_5C\bullet(OH)CH_3$ | $8.1.10^{+05}$ | 1.9 | 5670 | (148) |
| iso$C_4H_9OH+\bullet O\bullet=>\bullet OH+CH_3CH(CH_3)\bullet CHOH$ | $3.4.10^{+08}$ | 1.5 | 1000 | (149) |
| iso$C_4H_9OH+\bullet H=>H_2+CH_3CH(CH_3)\bullet CHOH$ | $4.8.10^{+09}$ | 1.5 | 3310 | (150) |
| iso$C_4H_9OH+\bullet OH=>H_2O+CH_3CH(CH_3)\bullet CHOH$ | $2.4.10^{+06}$ | 2.0 | -2200 | (151) |
| iso$C_4H_9OH+\bullet HO_2=>H_2O_2+CH_3CH(CH_3)\bullet CHOH$ | $2.8.10^{+04}$ | 2.7 | 14380 | (152) |
| iso$C_4H_9OH+\bullet CH_3=>CH_4+CH_3CH(CH_3)\bullet CHOH$ | $1.6.10^{+06}$ | 1.9 | 6840 | (153) |
| $C_4H_9OH$-1$+\bullet O\bullet=>\bullet OH+C_3H_7CH_2O\bullet$ | $1.7.10^{+13}$ | 0.0 | 7850 | (154) |
| $C_4H_9OH$-1$+\bullet H=>H_2+C_3H_7CH_2O\bullet$ | $9.5.10^{+06}$ | 2.0 | 7700 | (155) |
| $C_4H_9OH$-1$+\bullet OH=>H_2O+C_3H_7CH_2O\bullet$ | $8.9.10^{+05}$ | 2.0 | 450 | (156) |
| $C_4H_9OH$-1$+\bullet HO_2=>H_2O_2+C_3H_7CH_2O\bullet$ | $2.0.10^{+11}$ | 0.0 | 17000 | (157) |
| $C_4H_9OH$-1$+\bullet CH_3=>CH_4+C_3H_7CH_2O\bullet$ | $1.0.10^{-01}$ | 4.0 | 8200 | (158) |
| $C_4H_9OH$-2$+\bullet O\bullet=>\bullet OH+C_2H_5CH(O\bullet)CH_3$ | $1.7.10^{+13}$ | 0.0 | 7850 | (159) |
| $C_4H_9OH$-2$+\bullet H=>H_2+C_2H_5CH(O\bullet)CH_3$ | $9.5.10^{+06}$ | 2.0 | 7700 | (160) |
| $C_4H_9OH$-2$+\bullet OH=>H_2O+C_2H_5CH(O\bullet)CH_3$ | $8.9.10^{+05}$ | 2.0 | 450 | (161) |
| $C_4H_9OH$-2$+\bullet HO_2=>H_2O_2+C_2H_5CH(O\bullet)CH_3$ | $2.0.10^{+11}$ | 0.0 | 17000 | (162) |
| $C_4H_9OH$-2$+\bullet CH_3=>CH_4+C_2H_5CH(O\bullet)CH_3$ | $1.0.10^{-01}$ | 4.0 | 8200 | (163) |
| iso$C_4H_9OH+\bullet O\bullet=>\bullet OH+CH_3CH(CH_3)CH_2O\bullet$ | $1.7.10^{+13}$ | 0.0 | 7850 | (164) |
| iso$C_4H_9OH+\bullet H=>H_2+CH_3CH(CH_3)CH_2O\bullet$ | $9.5.10^{+06}$ | 2.0 | 7700 | (165) |
| iso$C_4H_9OH+\bullet OH=>H_2O+CH_3CH(CH_3)CH_2O\bullet$ | $8.9.10^{+05}$ | 2.0 | 450.0 | (166) |
| isoC$_4$H$_9$OH+•HO$_2$=>H$_2$O$_2$+CH$_3$CH(CH$_3$)CH$_2$O• | 2.0.10$^{+11}$ | 0.0 | 17000 | (167) |
| isoC$_4$H$_9$OH+•CH$_3$=>CH$_4$+CH$_3$CH(CH$_3$)CH$_2$O• | 1.0.10$^{-01}$ | 4.0 | 8200 | (168) |
| terC$_4$H$_9$OH+•O•=>•OH+C(CH$_3$)$_3$O• | 1.7.10$^{+13}$ | 0.0 | 7850 | (169) |
| terC$_4$H$_9$OH+•H=>H$_2$+C(CH$_3$)$_3$O• | 9.5.10$^{+06}$ | 2.0 | 7700 | (170) |
| terC$_4$H$_9$OH+•OH=>H$_2$O+C(CH$_3$)$_3$O• | 8.9.10$^{+05}$ | 2.0 | 450 | (171) |
| terC$_4$H$_9$OH+•HO$_2$=>H$_2$O$_2$+C(CH$_3$)$_3$O• | 2.0.10$^{+11}$ | 0.0 | 17000 | (172) |
| terC$_4$H$_9$OH+•CH$_3$=>CH$_4$+C(CH$_3$)$_3$O• | 1.0.10$^{-01}$ | 4.0 | 8200 | (173) |
| C$_4$H$_9$OH-1+•O•=>•OH+•CH$_2$C$_3$H$_6$OH | 5.1.10$^{+13}$ | 0.0 | 7850 | (174) |
| C$_4$H$_9$OH-1+•O•=>•OH+CH$_3$•CHC$_2$H$_4$OH | 2.6.10$^{+13}$ | 0.0 | 5200 | (175) |
| C$_4$H$_9$OH-1+•O•=>•OH+C$_2$H$_5$•CHCH$_2$OH | 2.6.10$^{+13}$ | 0.0 | 5200 | (176) |
| C$_4$H$_9$OH-1+•H=>H$_2$+•CH$_2$C$_3$H$_6$OH | 2.9.10$^{+07}$ | 2.0 | 7700 | (177) |
| C$_4$H$_9$OH-1+•H=>H$_2$+CH$_3$•CHC$_2$H$_4$OH | 9.0.10$^{+06}$ | 2.0 | 5000 | (178) |
| C$_4$H$_9$OH-1+•H=>H$_2$+C$_2$H$_5$•CHCH$_2$OH | 9.0.10$^{+06}$ | 2.0 | 5000 | (179) |
| C$_4$H$_9$OH-1+•OH=>H$_2$O+•CH$_2$C$_3$H$_6$OH | 2.7.10$^{+06}$ | 2.0 | 450 | (180) |
| C$_4$H$_9$OH-1+•OH=>H$_2$O+CH$_3$•CHC$_2$H$_4$OH | 2.6.10$^{+06}$ | 2.0 | -765 | (181) |
| C$_4$H$_9$OH-1+•OH=>H$_2$O+C$_2$H$_5$•CHCH$_2$OH | 2.6.10$^{+06}$ | 2.0 | -765 | (182) |
| C$_4$H$_9$OH-1+•HO$_2$=>H$_2$O$_2$+•CH$_2$C$_3$H$_6$OH | 6.0.10$^{+11}$ | 0.0 | 17000 | (183) |
| C$_4$H$_9$OH-1+•HO$_2$=>H$_2$O$_2$+CH$_3$•CHC$_2$H$_4$OH | 4.0.10$^{+11}$ | 0.0 | 15500 | (184) |
| C$_4$H$_9$OH-1+•HO$_2$=>H$_2$O$_2$+C$_2$H$_5$•CHCH$_2$OH | 4.0.10$^{+11}$ | 0.0 | 15500 | (185) |
| C$_4$H$_9$OH-1+•CH$_3$=>CH$_4$+•CH$_2$C$_3$H$_6$OH | 3.0.10$^{-01}$ | 4.0 | 8200 | (186) |
| C$_4$H$_9$OH-1+•CH$_3$=>CH$_4$+CH$_3$•CHC$_2$H$_4$OH | 2.0.10$^{+11}$ | 0.0 | 9600 | (187) |
| C$_4$H$_9$OH-1+•CH$_3$=>CH$_4$+C$_2$H$_5$•CHCH$_2$OH | 2.0.10$^{+11}$ | 0.0 | 9600 | (188) |
| C$_4$H$_9$OH-2+•O•=>•OH+C$_2$H$_5$CH(OH)•CH$_2$ | 5.1.10$^{+13}$ | 0.0 | 7850 | (189) |
| C$_4$H$_9$OH-2+•O•=>•OH+•CH$_2$CH$_2$CH(OH)CH$_3$ | 5.1.10$^{+13}$ | 0.0 | 7850 | (190) |
| C$_4$H$_9$OH-2+•O•=>•OH+CH$_3$•CHCH(OH)CH$_3$ | 2.6.10$^{+13}$ | 0.0 | 5200 | (191) |
| C$_4$H$_9$OH-2+•H=>H$_2$+C$_2$H$_5$CH(OH)•CH$_2$ | 2.9.10$^{+07}$ | 2.0 | 7700 | (192) |
| C$_4$H$_9$OH-2+•H=>H$_2$+•CH$_2$CH$_2$CH(OH)CH$_3$ | 2.9.10$^{+07}$ | 2.0 | 7700 | (193) |
| C$_4$H$_9$OH-2+•H=>H$_2$+CH$_3$•CHCH(OH)CH$_3$ | 9.0.10$^{+06}$ | 2.0 | 5000 | (194) |
| C$_4$H$_9$OH-2+•OH=>H$_2$O+C$_2$H$_5$CH(OH)•CH$_2$ | 2.7.10$^{+06}$ | 2.0 | 450 | (195) |
| C$_4$H$_9$OH-2+•OH=>H$_2$O+•CH$_2$CH$_2$CH(OH)CH$_3$ | 2.7.10$^{+06}$ | 2.0 | 450 | (196) |
| C$_4$H$_9$OH-2+•OH=>H$_2$O+CH$_3$•CHCH(OH)CH$_3$ | 2.6.10$^{+06}$ | 2.0 | -765 | (197) |
| C$_4$H$_9$OH-2+•HO$_2$=>H$_2$O$_2$+C$_2$H$_5$CH(OH)•CH$_2$ | 6.0.10$^{+11}$ | 0.0 | 17000 | (198) |
| C$_4$H$_9$OH-2+•HO$_2$=>H$_2$O$_2$+•CH$_2$CH$_2$CH(OH)CH$_3$ | 6.0.10$^{+11}$ | 0.0 | 17000 | (199) |
| C$_4$H$_9$OH-2+•HO$_2$=>H$_2$O$_2$+CH$_3$•CHCH(OH)CH$_3$ | 4.0.10$^{+11}$ | 0.0 | 15500 | (200) |
| C$_4$H$_9$OH-2+•CH$_3$=>CH$_4$+C$_2$H$_5$CH(OH)•CH$_2$ | 3.0.10$^{-01}$ | 4.0 | 8200 | (201) |
| C$_4$H$_9$OH-2+•CH$_3$=>CH$_4$+•CH$_2$CH$_2$CH(OH)CH$_3$ | 3.0.10$^{-01}$ | 4.0 | 8200 | (202) |
| C$_4$H$_9$OH-2+•CH$_3$=>CH$_4$+CH$_3$•CHCH(OH)CH$_3$ | 2.0.10$^{+11}$ | 0.0 | 9600 | (203) |
| isoC$_4$H$_9$OH+•O•=>•OH+CH$_3$C• (CH$_3$)CH$_2$OH | 1.0.10$^{+13}$ | 0.0 | 3280 | (204) |
| isoC$_4$H$_9$OH+•O•=>•OH+•CH$_2$CH(CH$_3$)CH$_2$OH | 1.0.10$^{+14}$ | 0.0 | 7850 | (205) |
| isoC$_4$H$_9$OH+•H=>H$_2$+CH$_3$C•(CH$_3$)CH$_2$OH | 4.2.10$^{+06}$ | 2.0 | 2400 | (206) |
| isoC$_4$H$_9$OH+•H=>H$_2$+•CH$_2$CH(CH$_3$)CH$_2$OH | 5.7.10$^{+07}$ | 2.0 | 7700 | (207) |
| isoC$_4$H$_9$OH+•OH=>H$_2$O+CH$_3$C•(CH$_3$)CH$_2$OH | 1.1.10$^{+06}$ | 2.0 | -1865 | (208) |
| isoC$_4$H$_9$OH+•OH=>H$_2$O+•CH$_2$CH(CH$_3$)CH$_2$OH | 5.4.10$^{+06}$ | 2.0 | 450 | (209) |
| isoC$_4$H$_9$OH+•HO$_2$=>H$_2$O$_2$+CH$_3$C•(CH$_3$)CH$_2$OH | 2.0.10$^{+11}$ | 0.0 | 14000 | (210) |
| isoC$_4$H$_9$OH+•HO$_2$=>H$_2$O$_2$+•CH$_2$CH(CH$_3$)CH$_2$OH | 1.2.10$^{+12}$ | 0.0 | 17000 | (211) |
| isoC$_4$H$_9$OH+•CH$_3$=>CH$_4$+CH$_3$C•(CH$_3$)CH$_2$OH | 1.0.10$^{+11}$ | 0.0 | 7900 | (212) |
| isoC$_4$H$_9$OH+•CH$_3$=>CH$_4$+•CH$_2$CH(CH$_3$)CH$_2$OH | 6.0.10$^{-01}$ | 4.0 | 8200 | (213) |
| terC$_4$H$_9$OH+•O•=>•OH+•CH$_2$C(CH$_3$)$_2$OH | 1.5.10$^{+14}$ | 0.0 | 7850 | (214) |
| terC$_4$H$_9$OH+•H=>H$_2$+•CH$_2$C(CH$_3$)$_2$OH | 8.6.10$^{+07}$ | 2.0 | 7700 | (215) |
| terC$_4$H$_9$OH+•OH=>H$_2$O+•CH$_2$C(CH$_3$)$_2$OH | 8.1.10$^{+06}$ | 2.0 | 450 | (216) |
| terC$_4$H$_9$OH+•HO$_2$=>H$_2$O$_2$+•CH$_2$C(CH$_3$)$_2$OH | 1.8.10$^{+12}$ | 0.0 | 17000 | (217) |
| terC$_4$H$_9$OH+•CH$_3$=>CH$_4$+•CH$_2$C(CH$_3$)$_2$OH | 9.0.10$^{-01}$ | 4.0 | 8200 | (218) |
| **Combinations :** | | | | |
| •H+•C$_3$H$_7$-2=>C$_3$H$_8$ | 8.3.10$^{+12}$ | 0.0 | 0 | (219) |
| •H+•tC$_4$H$_9$=>C$_4$H$_{10}$ | 8.3.10$^{+12}$ | 0.0 | 0 | (220) |
| •OH+•C$_3$H$_7$-2=>C$_3$H$_7$OH | 5.9.10+$^{12}$ | 0.0 | 0 | (221) |
| •HO$_2$+•C$_3$H$_7$-2=>C$_3$H$_7$OOH | 4.8.10$^{+12}$ | 0.0 | 0 | (222) |
| •HO$_2$+•tC$_4$H$_9$=>C$_4$H$_9$OOH | 4.5.10$^{+12}$ | 0.0 | 0 | (223) |
| •CH$_3$+•C$_3$H$_7$-2=>C$_4$H$_{10}$ | 1.5.10$^{+13}$ | 0.0 | 0 | (224) |
| •CH$_3$+•tC$_4$H$_9$=>C$_5$H$_{12}$ | 1.5.10$^{+13}$ | 0.0 | 0 | (225) |
| •HCO+•C$_3$H$_7$-2=>C$_4$H$_8$O-A | 5.2.10$^{+12}$ | 0.0 | 0 | (226) |
| •HCO+•tC$_4$H$_9$=>C$_5$H$_{10}$O-A | 4.9.10$^{+12}$ | 0.0 | 0 | (227) |
| •CH$_2$OH+•tC$_4$H$_9$=>C$_5$H$_{12}$O-L | 4.8.10$^{+12}$ | 0.0 | 0 | (228) |
| $CH_3O\bullet + \bullet C_3H_7\text{-}2 \Rightarrow C_4H_{10}O\text{-}E$ [d] | $4.9.10^{+12}$ | 0.0 | 0 | (229) |
| $CH_3O\bullet + \bullet tC_4H_9 \Rightarrow C_5H_{12}O\text{-}E$ | $4.6.10^{+12}$ | 0.0 | 0 | (230) |
| $CH_3OO\bullet + \bullet C_3H_7\text{-}2 \Rightarrow C_4H_{10}O_2\text{-}U$ [e] | $4.4.10^{+12}$ | 0.0 | 0 | (231) |
| $CH_3OO\bullet + \bullet tC_4H_9 \Rightarrow C_5H_{12}O_2\text{-}U$ | $4.1.10^{+12}$ | 0.0 | 0 | (232) |
| $\bullet C_2H_5 + \bullet C_3H_7\text{-}2 \Rightarrow C_5H_{12}$ | $5.2.10^{+12}$ | 0.0 | 0 | (233) |
| $\bullet C_2H_5 + \bullet tC_4H_9 \Rightarrow C_6H_{14}$ | $4.9.10^{+12}$ | 0.0 | 0 | (234) |
| $\bullet C_3H_7\text{-}2 + \bullet C_3H_7\text{-}2 \Rightarrow C_6H_{14}$ | $2.3.10^{+12}$ | 0.0 | 0 | (235) |
| $\bullet C_3H_7\text{-}2 + \bullet tC_4H_9 \Rightarrow C_7H_{16}$ | $4.3.10^{+12}$ | 0.0 | 0 | (236) |
| $\bullet tC_4H_9 + \bullet tC_4H_9 \Rightarrow C_8H_{18}$ | $2.0.10^{+12}$ | 0.0 | 0 | (237) |

Notes :

[a] : L indicates that the species is bearing an alcohol function.

[b] : Y indicates that the species is an unsaturated one.

[c] : A indicates that the species is bearing an aldehyde function.

[d] : E indicates that the species is bearing an ether function.

[e] : U indicates that the species is bearing an -O-O• function.

**FIGURE CAPTIONS**

Figure 1: Example butanol ignition delay time measurement (pressure and OH* emission).

Figure 2: Raw 1-butanol ignition time measurements.

Figure 3: Correlated ignition times for all four butanol isomers.

Figure 4: Ignition time measurements for all four butanol isomers for a mixture composition of 1% butanol / 6% $O_2$ / Ar ($\Phi$ = 1.0) and reflected shock pressures near 1 bar.

Figure 5: Comparison of the current ignition time measurements for 1-butanol with the measurements of Zhukov *et al.* (13); all data scaled to a common condition: 1% butanol, $\Phi$ = 1.0, and 1 bar.

Figure 6: Example of intramolecular dehydration in the case of 1-butanol.

Figure 7: Comparison between simulations and experimental results for the ignition delay times of the four isomers of butanol for stoichiometric mixtures containing 1% butanol.

Figure 8: Comparison between simulations and experimental results for the ignition delay times of 1-butanol for (a) mixtures containing 0.5% 1-butanol for three different equivalence ratios and (b) for mixtures with an equivalence ratio of 0.5 for two different concentrations of 1-butanol.

Figure 9: Comparison between simulations and experimental results for the ignition delay times of 2-butanol for (a) mixtures containing 0.5% 2-butanol for three different equivalence ratios and (b) for mixtures with an equivalence ratio of 0.5 for two different concentrations of 2-butanol.

Figure 10: Comparison between simulations and experimental results for the ignition delay times of iso-butanol for (a) mixtures containing 0.5% iso-butanol for three different equivalence ratios and (b) for mixtures with an equivalence ratio of 0.5 for two different concentrations of iso-butanol.

Figure 11: Comparison between simulations and experimental results for the ignition delay times of tert-butanol for (a) mixtures containing 0.5% tert-butanol for three different equivalence ratios and (b) for mixtures with an equivalence ratio of 0.5 for two different concentrations of tert-butanol.

Figure 12: Reaction pathways analyses for the four isomers of butanol performed at 1450 K at atmospheric pressure for a stoichiometric mixtures containing 1% butanol and for 50% butanol conversion. The size of the arrows is proportional to the relative reaction flux. X• is H•, OH•, $HO_2$•, $CH_3$• or •O• radicals and XH is $H_2$, $H_2O$, $H_2O_2$, $CH_4$ molecules or OH• radicals.

Figure 13: Time evolution of the mole fraction of the primary species formed during the oxidation of the four butanol isomers under the conditions of figure 12.

Figure 14: Sensitivity analysis for the main consumption reactions for the four butanol isomers under the conditions of figure 12. Relative variations have been obtained by multiplying the rate constant of each generic reaction by a factor 10.

FIGURE 1

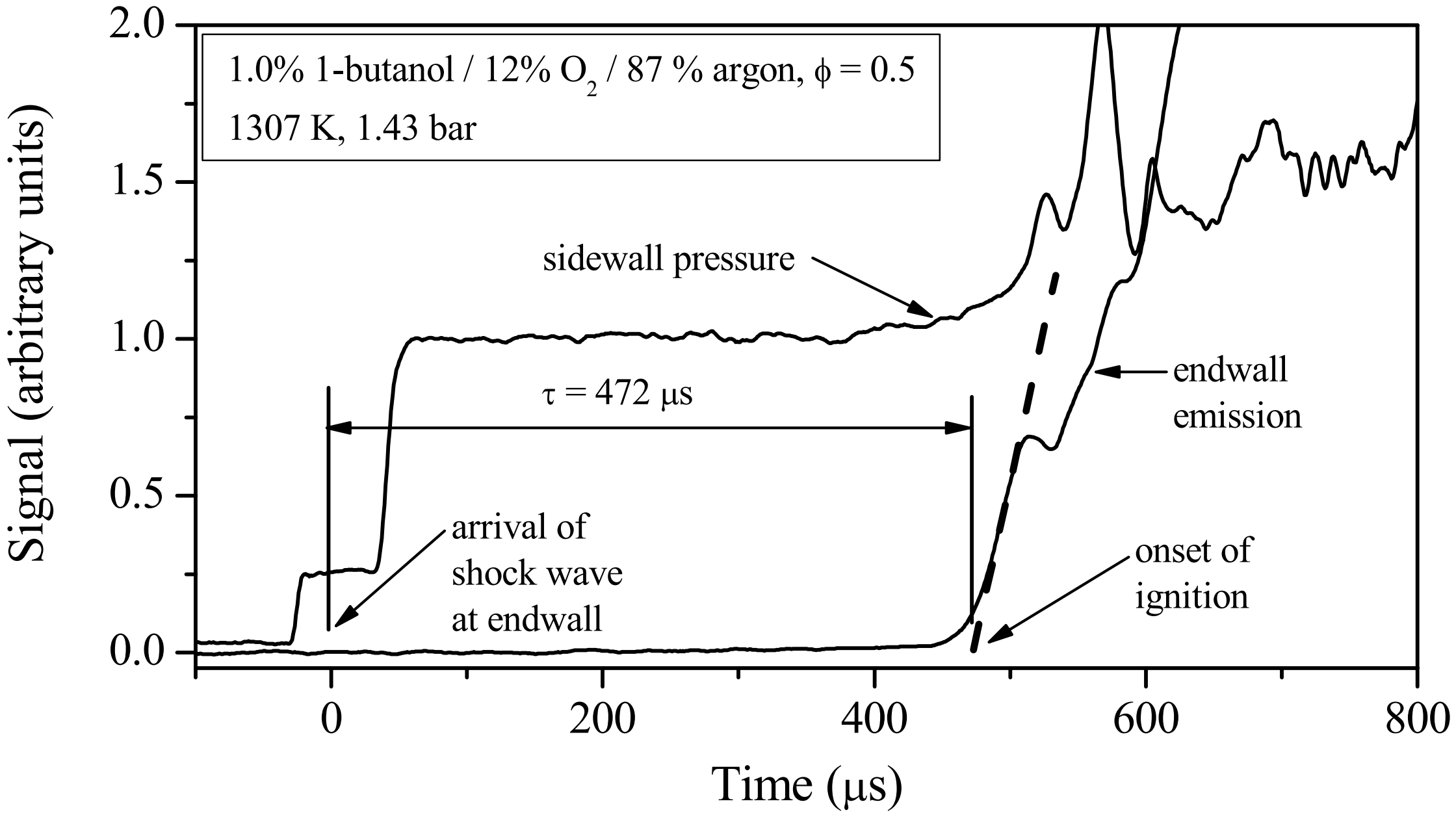

1.0% 1-butanol / 12% O2 / 87 % argon, ϕ = 0.5
1307 K, 1.43 bar
sidewall pressure
τ = 472 μs
endwall emission
arrival of shock wave at endwall
onset of ignition
Signal (arbitrary units)
Time (μs)
2.0
1.5
1.0
0.5
0.0
0
200
400
600
800

FIGURE 2

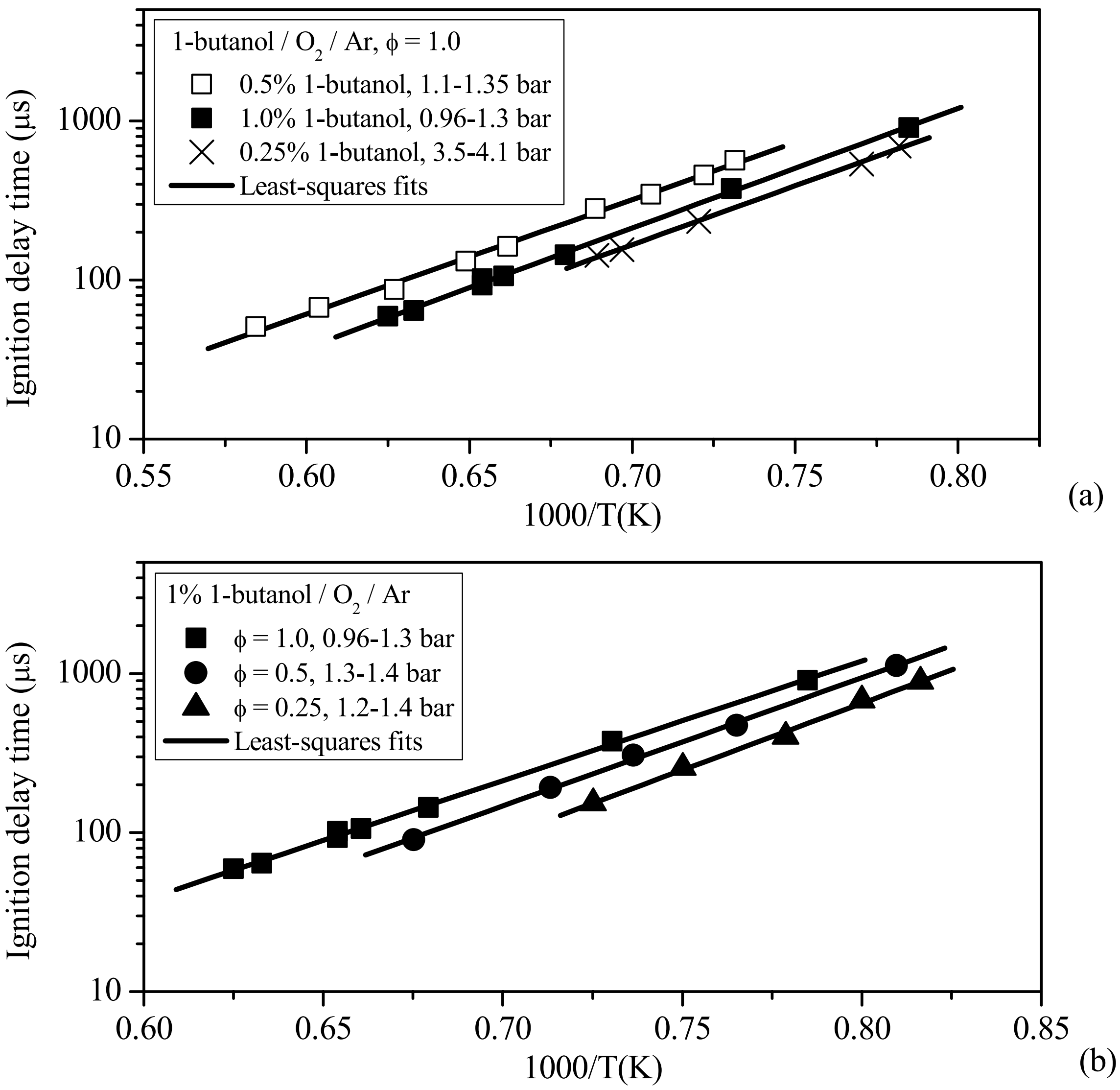
1-butanol / O2 / Ar, ϕ = 1.0
0.5% 1-butanol, 1.1-1.35 bar
1.0% 1-butanol, 0.96-1.3 bar
0.25% 1-butanol, 3.5-4.1 bar
Least-squares fits
Ignition delay time (µs)
10
100
1000
0.55
0.60
0.65
0.70
0.75
0.80
1000/T(K)
(a)
1% 1-butanol / O2 / Ar
ϕ = 1.0, 0.96-1.3 bar
ϕ = 0.5, 1.3-1.4 bar
ϕ = 0.25, 1.2-1.4 bar
Least-squares fits
Ignition delay time (µs)
10
100
1000
0.60
0.65
0.70
0.75
0.80
0.85
1000/T(K)
(b)

FIGURE 3

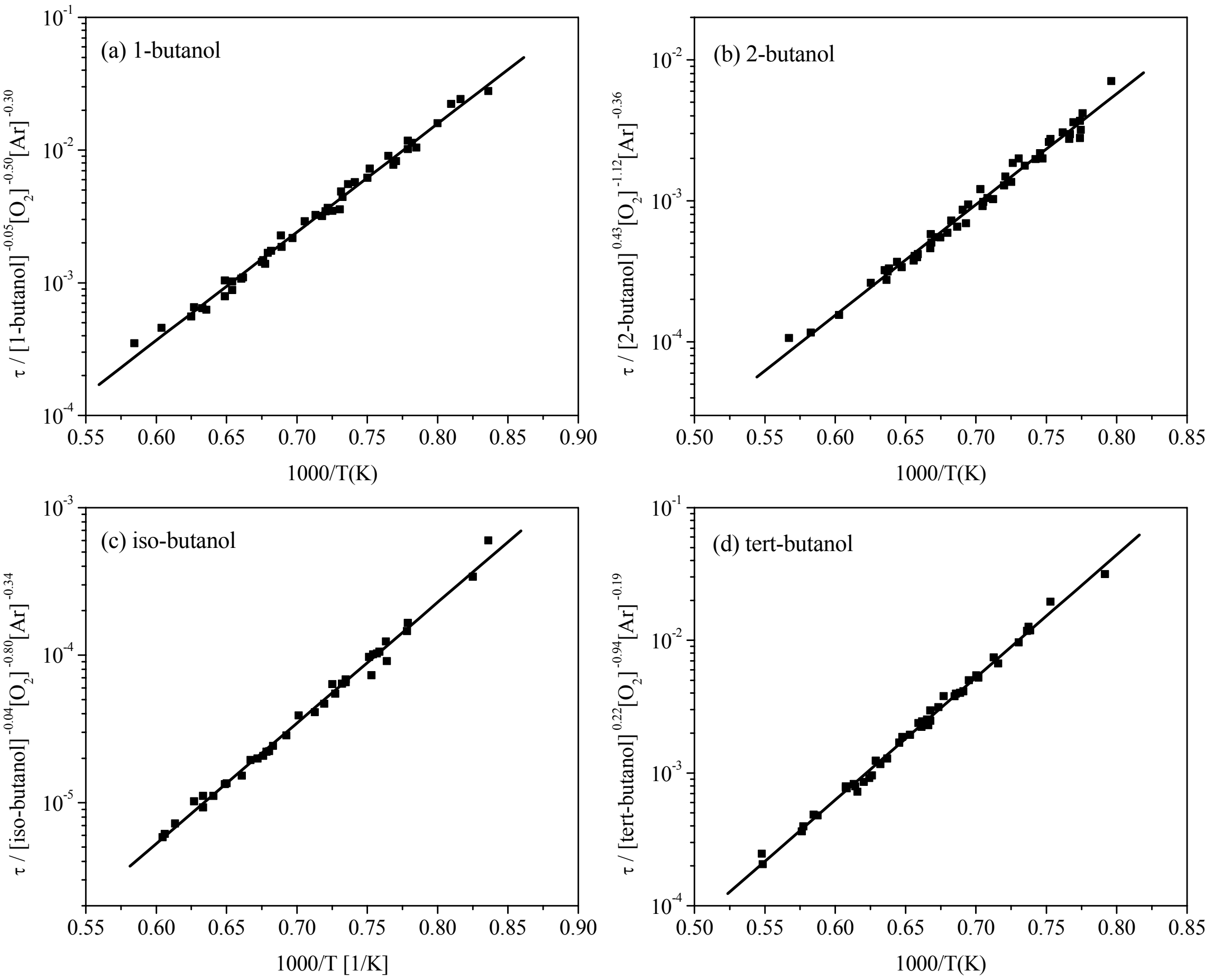

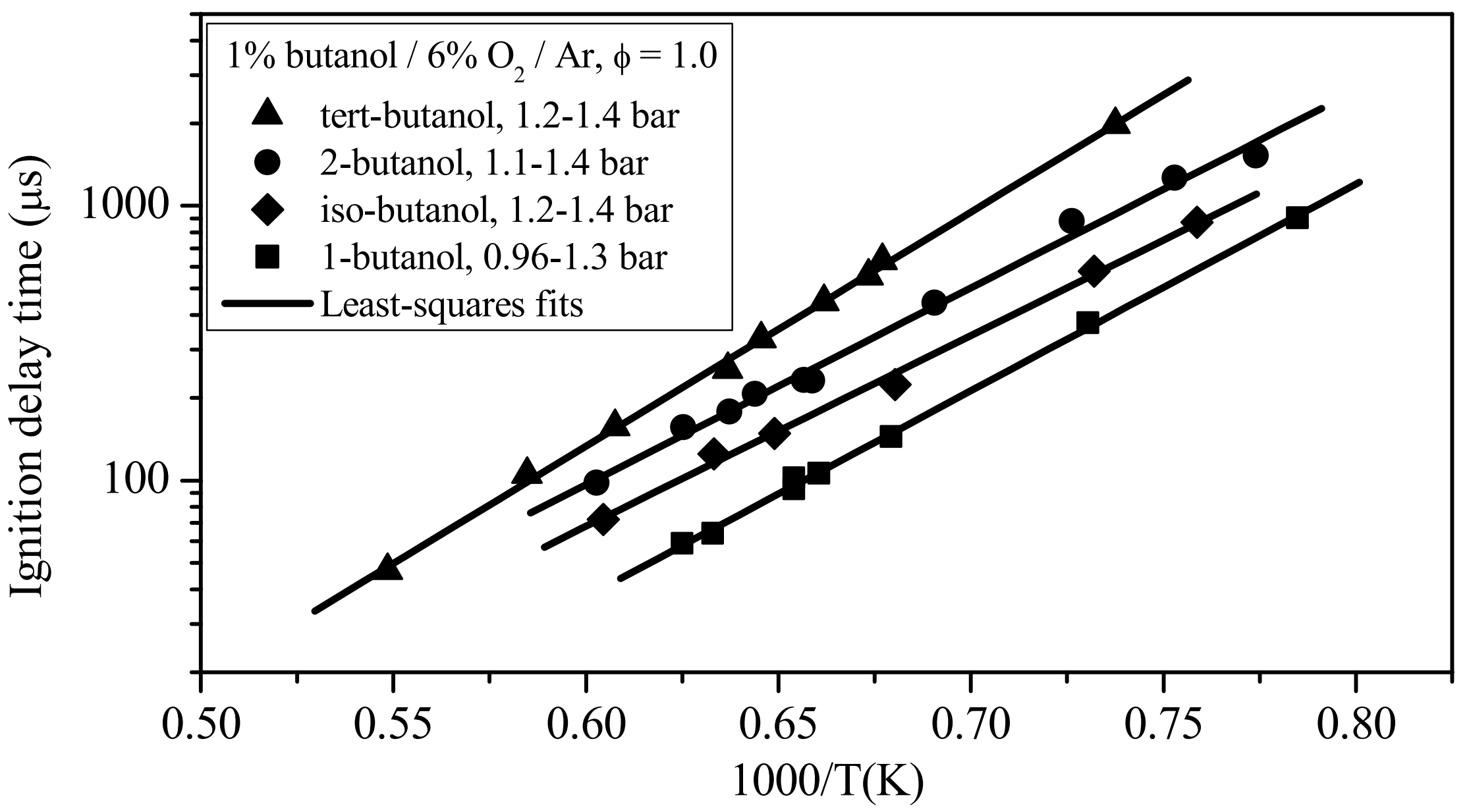
1% butanol / 6% $O_2$ / Ar, ϕ = 1.0
tert-butanol, 1.2-1.4 bar
2-butanol, 1.1-1.4 bar
iso-butanol, 1.2-1.4 bar
1-butanol, 0.96-1.3 bar
Least-squares fits
Ignition delay time (μs)
1000
100
0.50
0.55
0.60
0.65
0.70
0.75
0.80
1000/T(K)

FIGURE 5

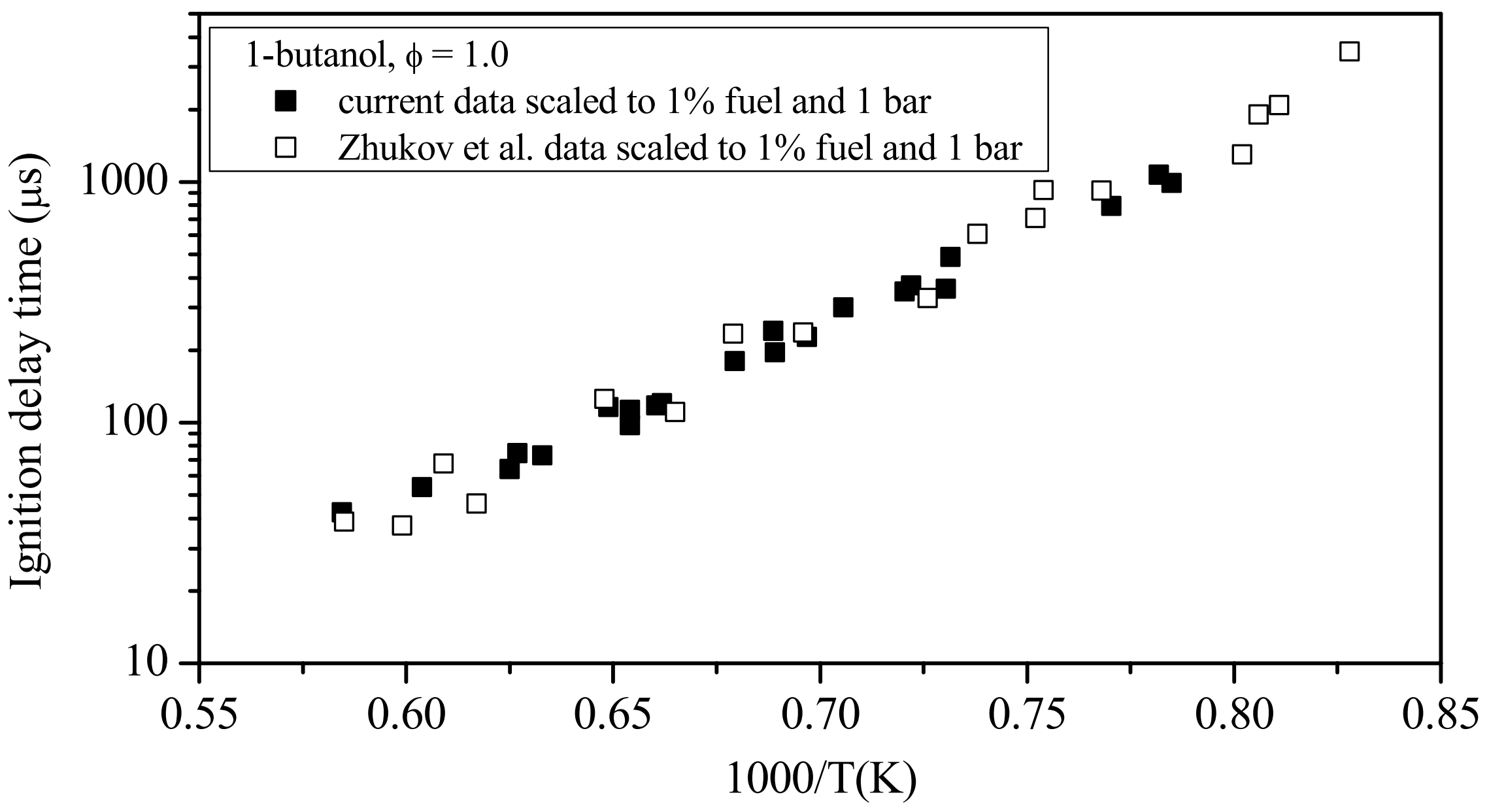

1-butanol, ϕ = 1.0
current data scaled to 1% fuel and 1 bar
Zhukov et al. data scaled to 1% fuel and 1 bar
Ignition delay time (μs)
1000/T(K)
10
100
1000
0.55
0.60
0.65
0.70
0.75
0.80
0.85

FIGURE 6

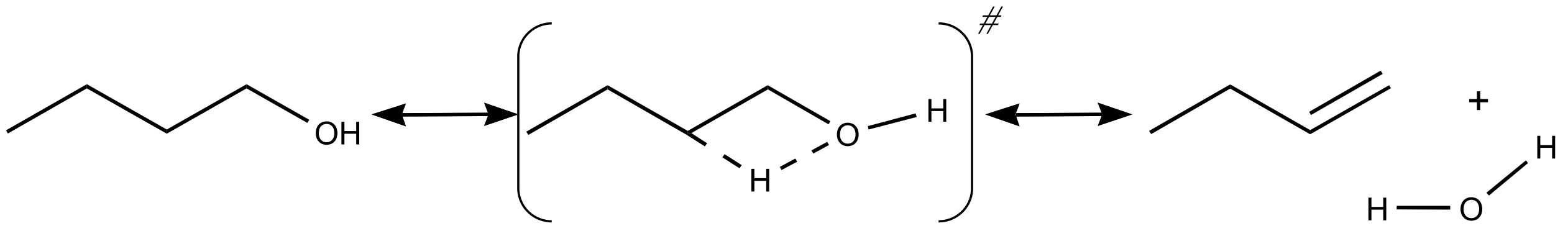

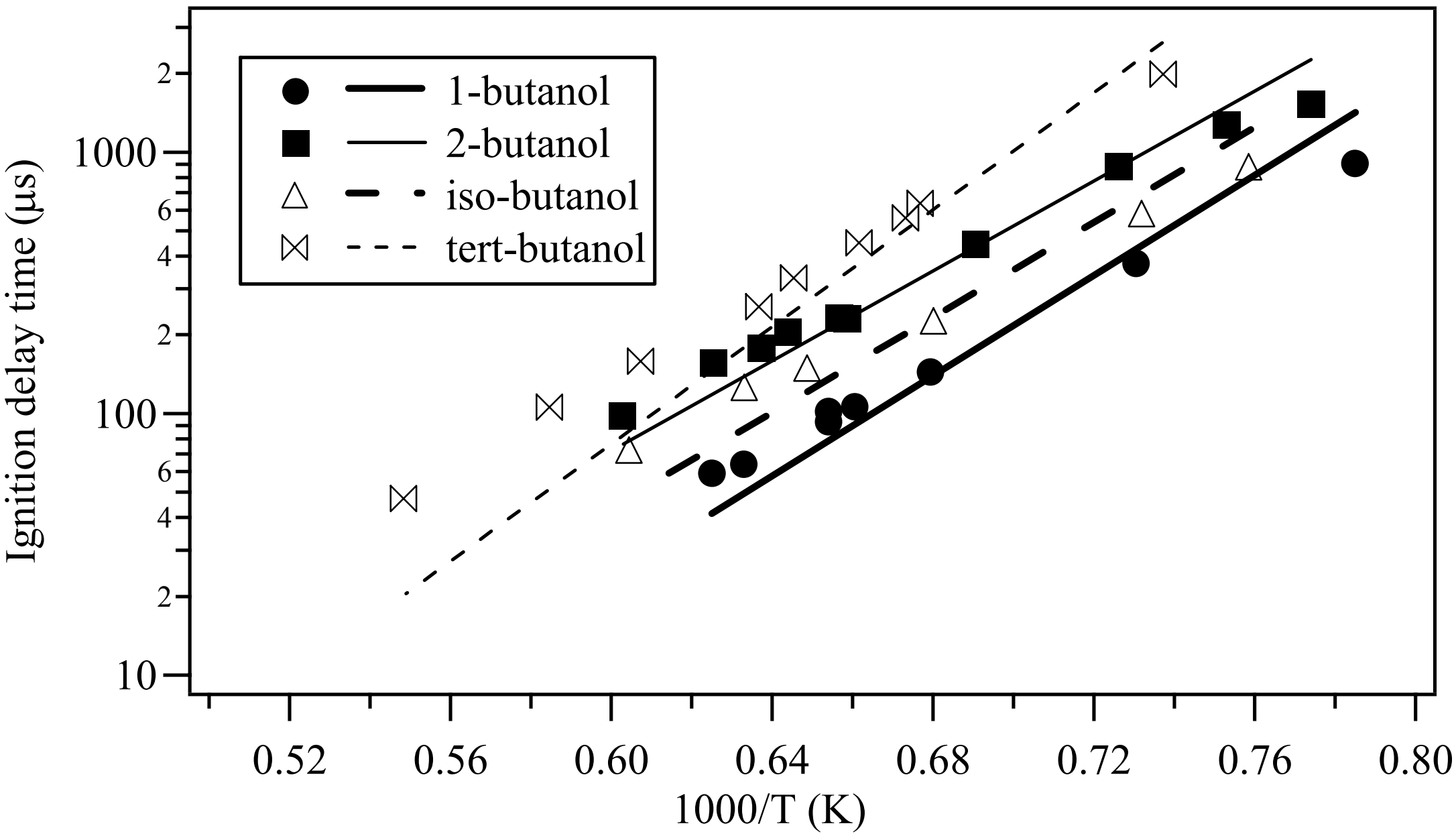

FIGURE 8

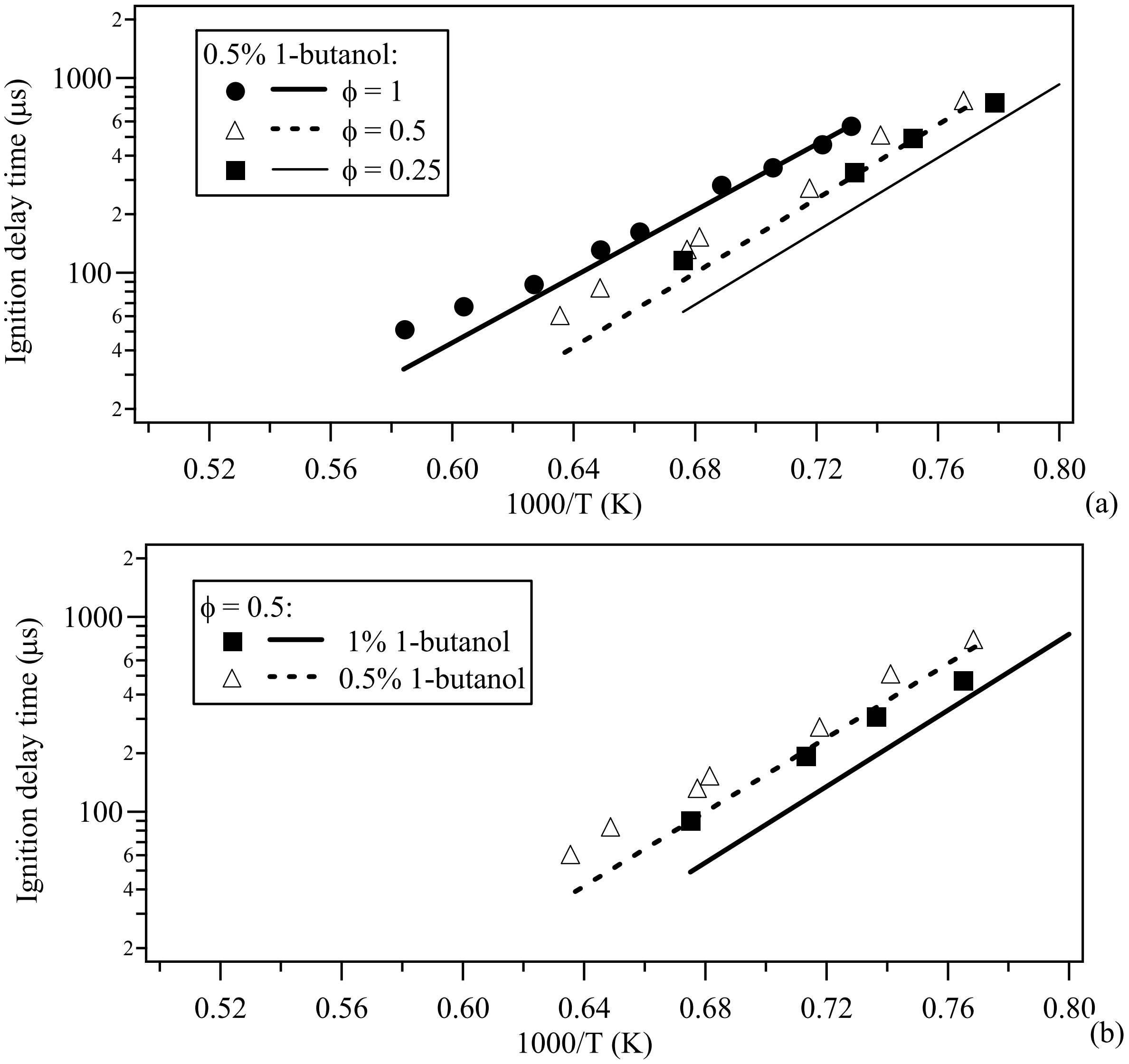

FIGURE 9

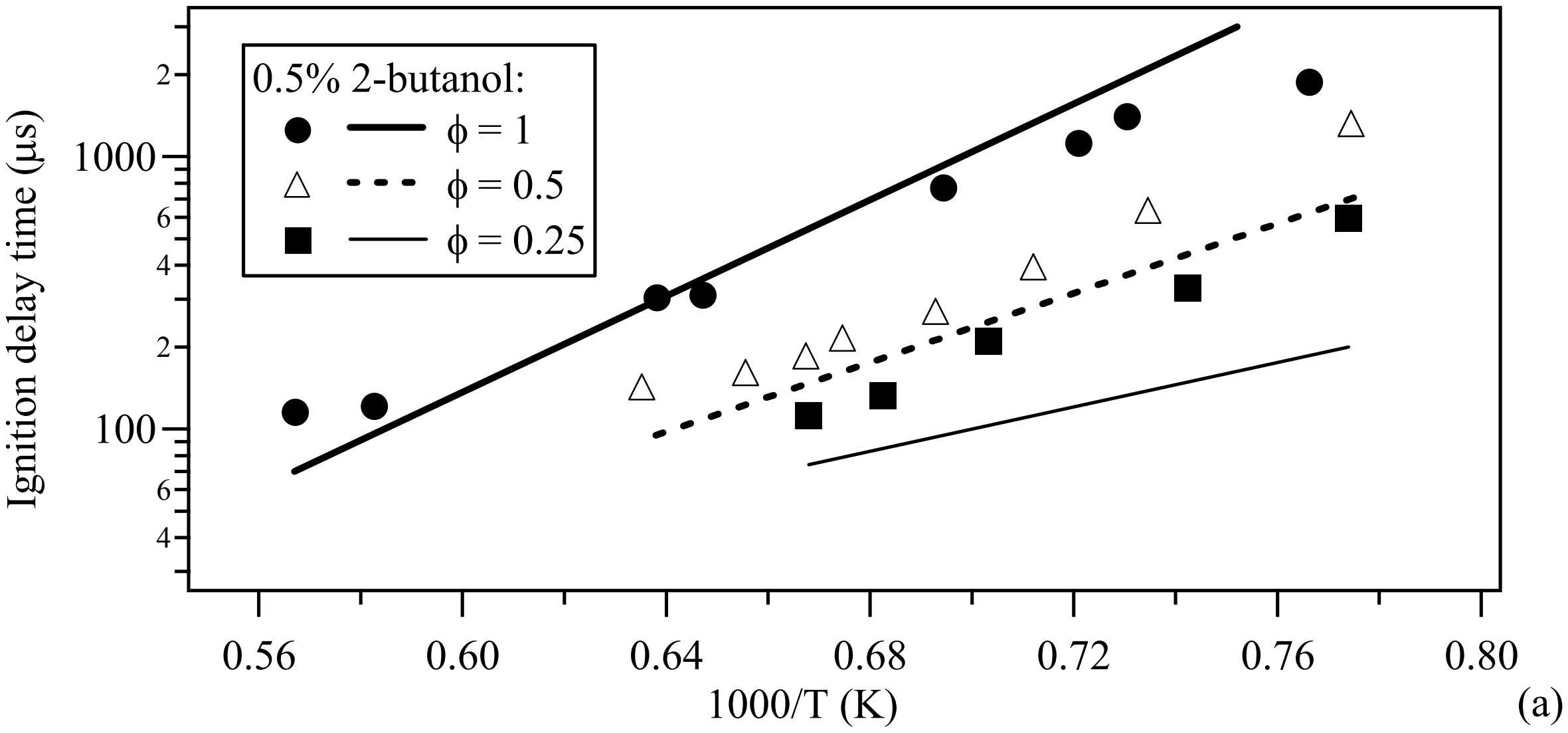

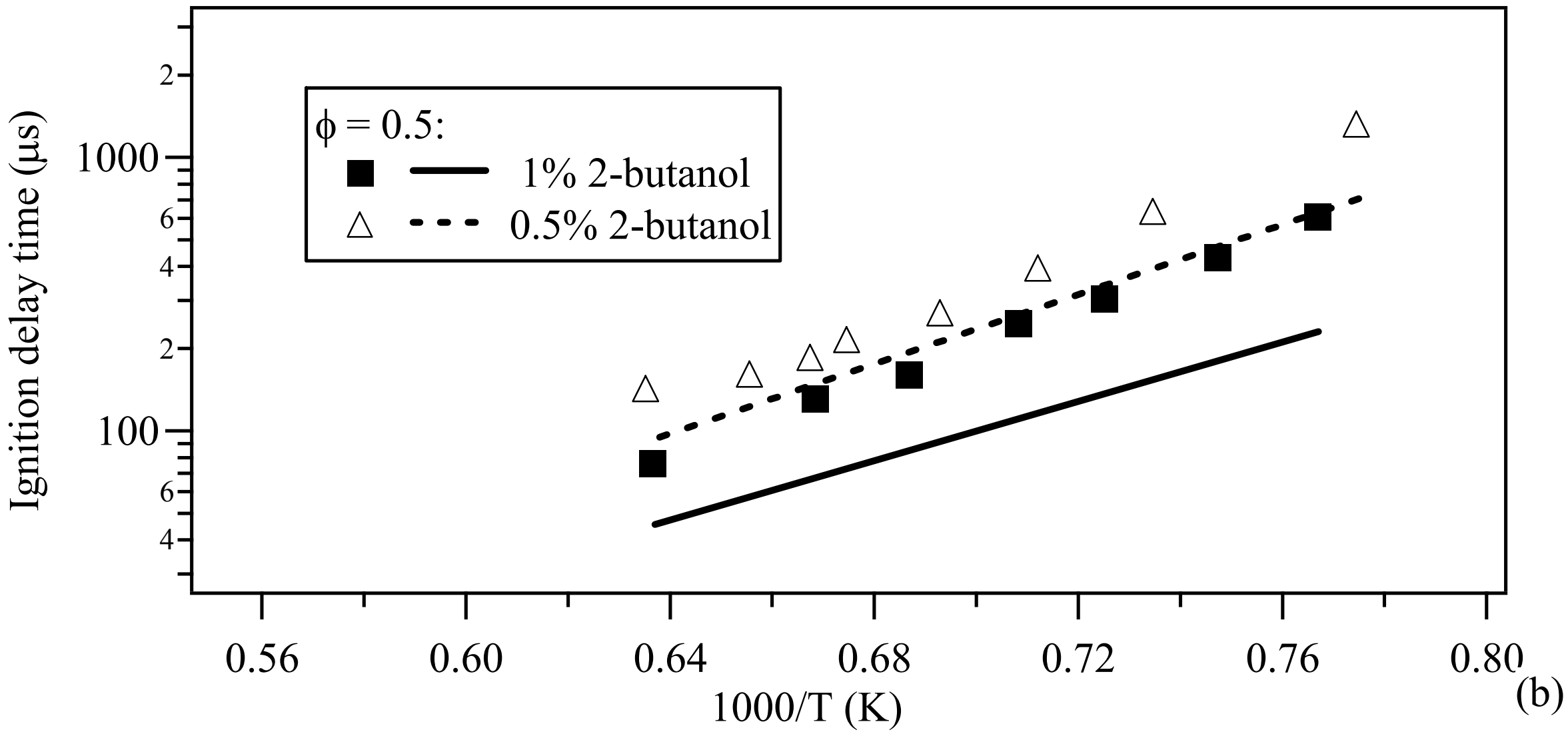

FIGURE 10

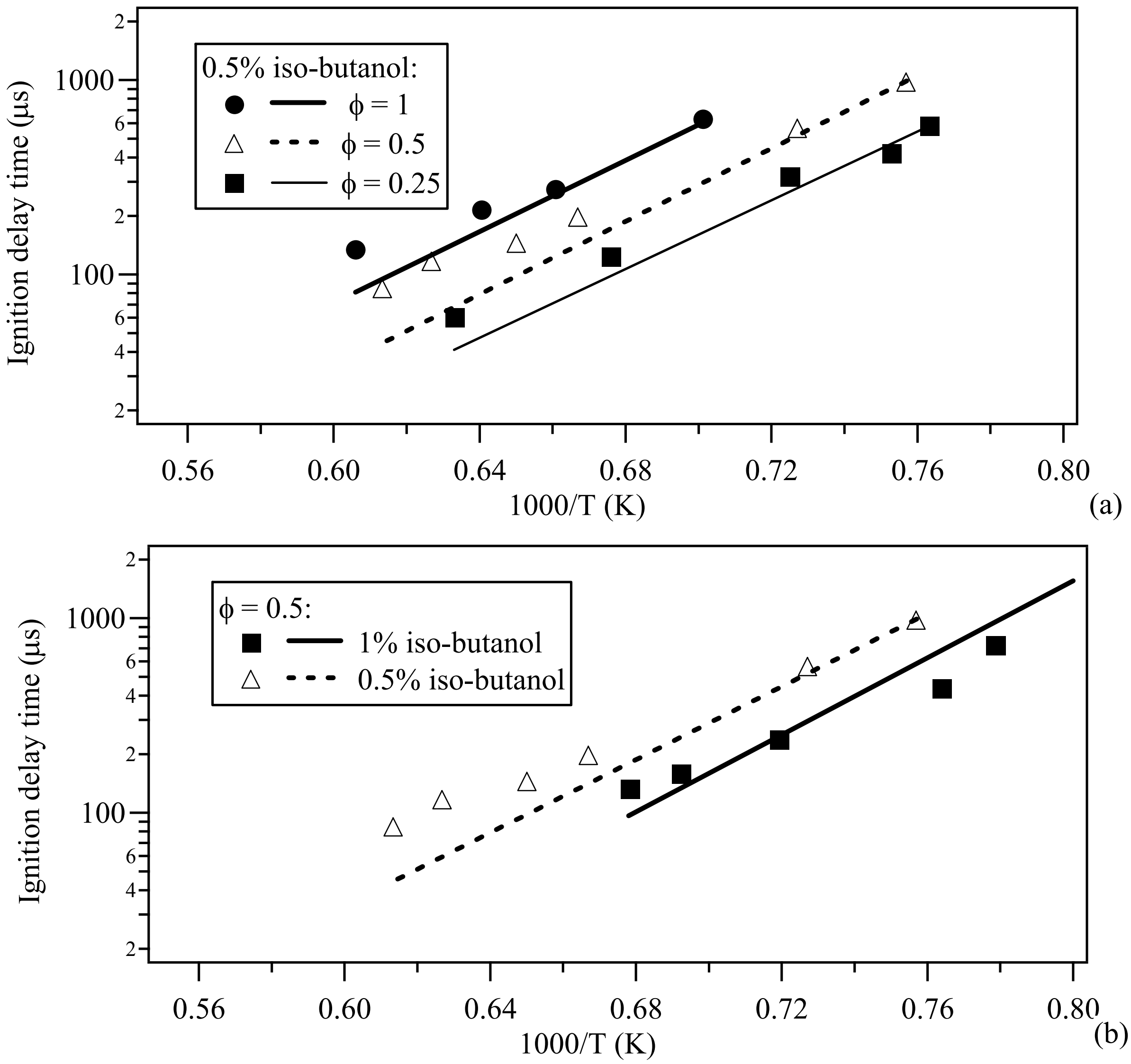

FIGURE 11

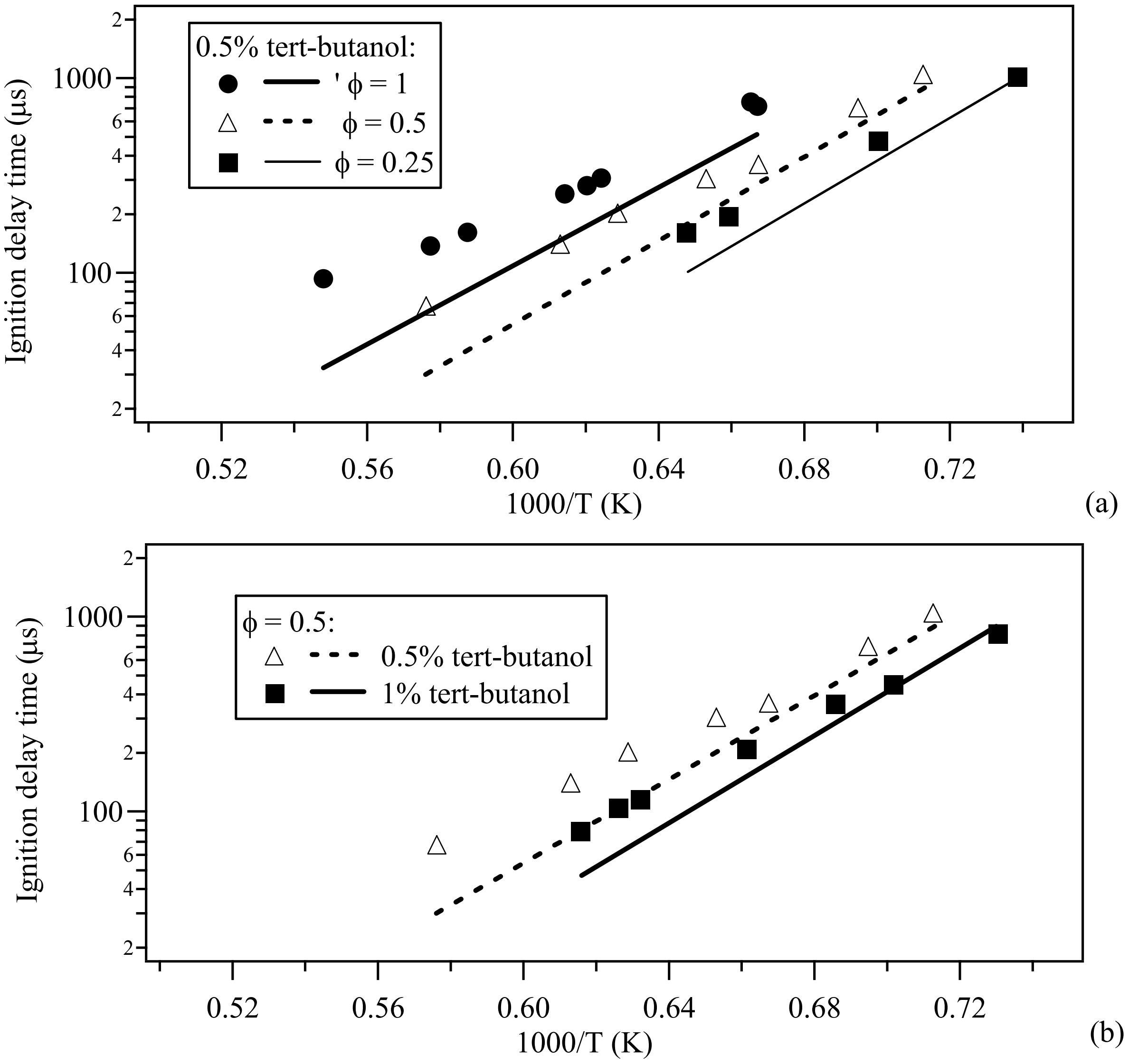

FIGURE 12

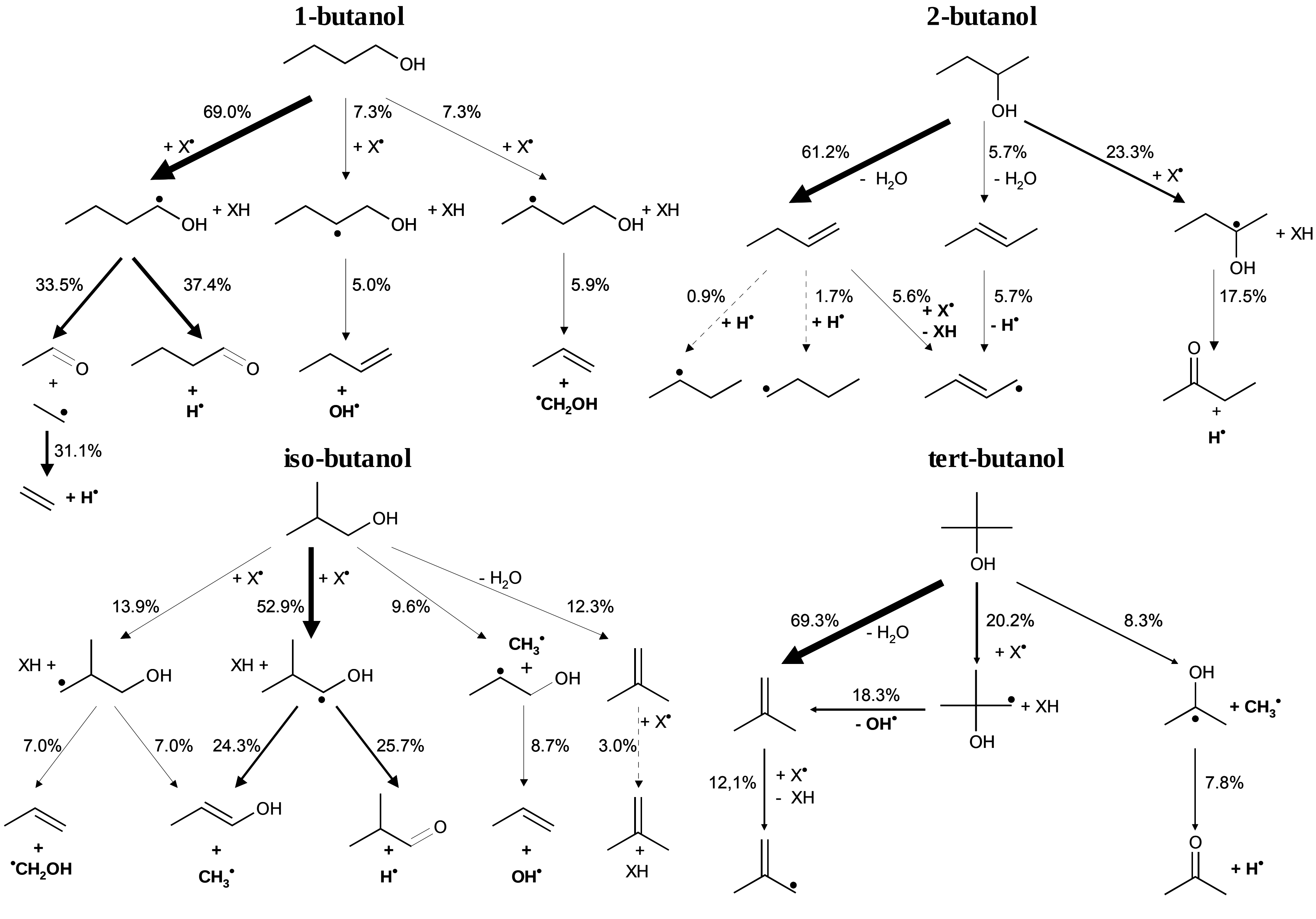

FIGURE 13

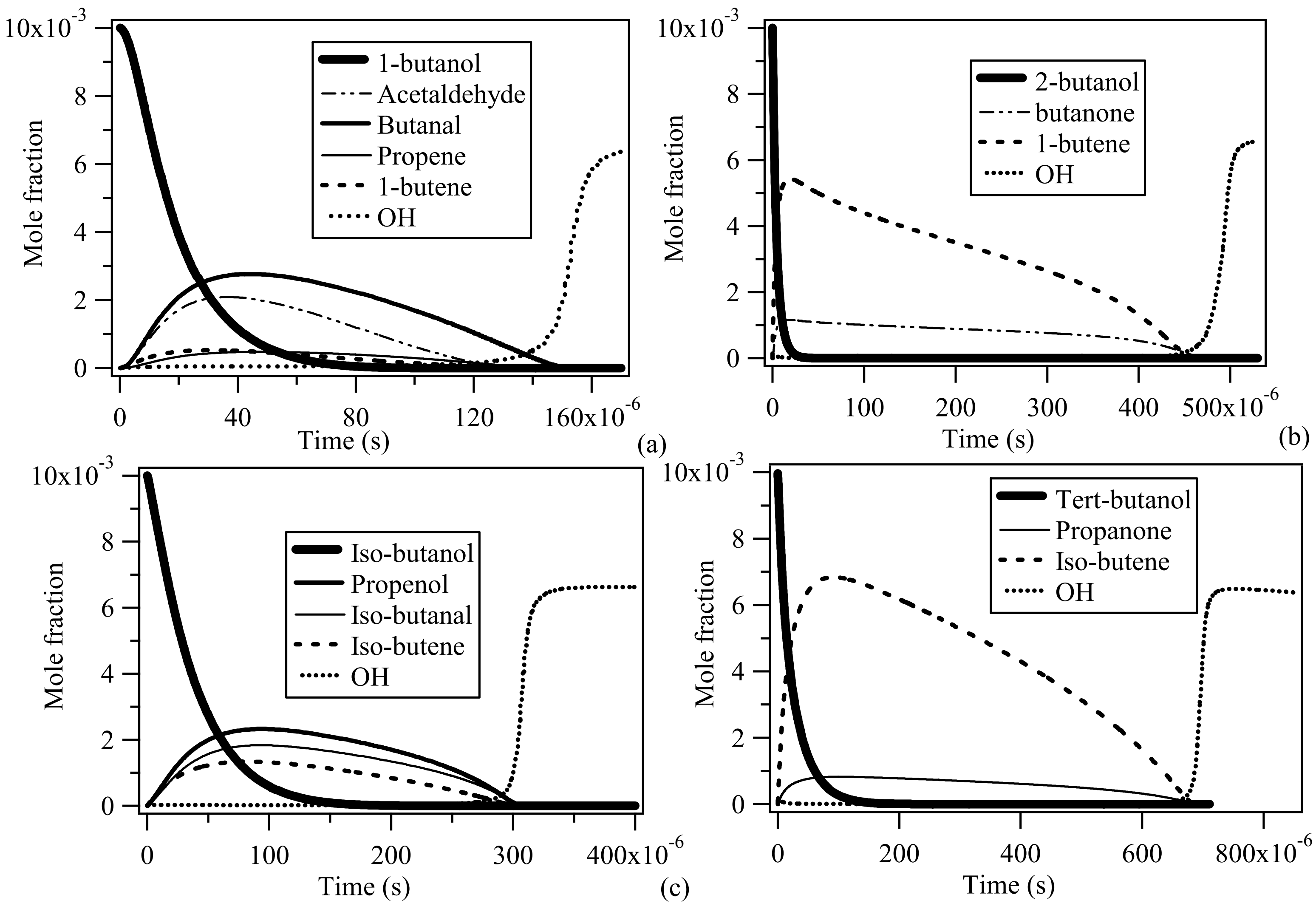

FIGURE 14

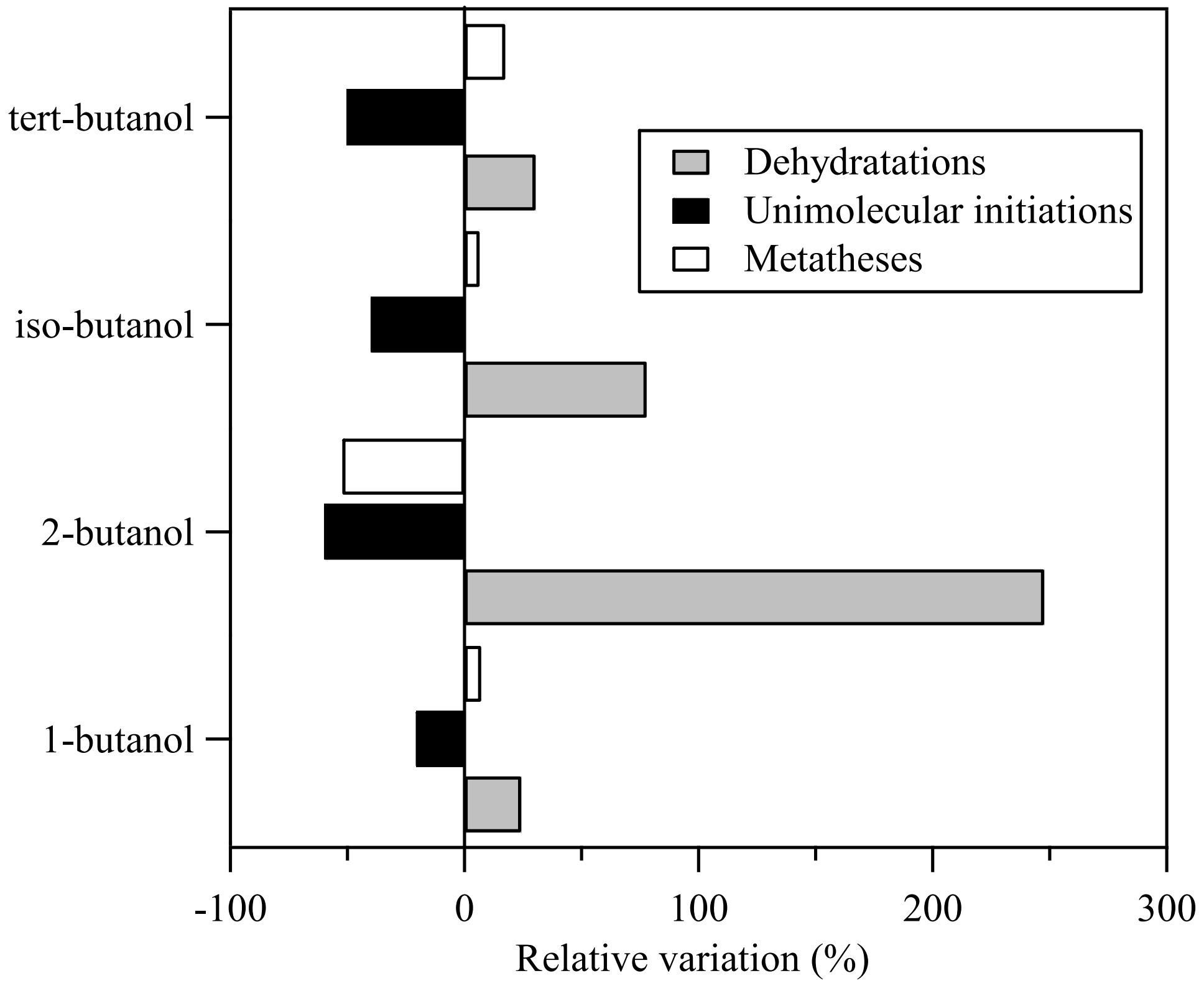